\documentclass[aps,prd,reprint,longbibliography,titlepage,nofootinbib]{revtex4-1}

\usepackage{amsthm}
\usepackage{amssymb}   
\usepackage{mathtools} 
\usepackage{hyperref}
\hypersetup{colorlinks=true,linkcolor=blue,urlcolor=blue,citecolor=blue}
\usepackage{accents}
\usepackage{tensor}
\usepackage[cal=boondox]{mathalfa}
\usepackage{lipsum}
\usepackage{relsize}
\usepackage{color}
\usepackage{comment}
\usepackage{nameref}
\usepackage[T1]{fontenc}

\newcommand{\uac}[1]{\underaccent{\tilde}{#1}}

\begin{document}
	
	\title{Canonical analysis of $n$-dimensional Palatini action without second-class constraints}

	\author{Merced Montesinos\href{https://orcid.org/0000-0002-4936-9170} {\includegraphics[scale=0.05]{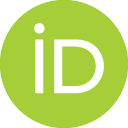}}}
	\email{merced@fis.cinvestav.mx}
	\author{Ricardo Escobedo\href{https://orcid.org/0000-0001-5815-4748} {\includegraphics[scale=0.05]{ORCIDiD_icon128x128.png}}}
	\email{rescobedo@fis.cinvestav.mx}
	\author{Jorge Romero\href{https://orcid.org/0000-0001-8258-6647} {\includegraphics[scale=0.05]{ORCIDiD_icon128x128.png}}}
	\email{ljromero@fis.cinvestav.mx}
	
	\affiliation{Departamento de F\'{i}sica, Cinvestav, Avenida Instituto Polit\'{e}cnico Nacional 2508,\\
		San Pedro Zacatenco, 07360 Gustavo A. Madero, Ciudad de M\'exico, Mexico}
	
	\author{Mariano Celada\href{https://orcid.org/0000-0002-3519-4736} {\includegraphics[scale=0.05]{ORCIDiD_icon128x128.png}}}
	\email[]{mcelada@matmor.unam.mx}
	\affiliation{Centro de Ciencias Matem\'{a}ticas, Universidad Nacional Aut\'{o}noma de M\'{e}xico,\\
	UNAM-Campus Morelia, Apartado Postal 61-3, Morelia, Michoac\'{a}n 58090, Mexico}
	
	\date{\today}
	
	\begin{abstract}
		We carry out the canonical analysis of the $n$-dimensional Palatini action with or without a cosmological constant $(n\geq3)$ introducing neither second-class constraints nor resorting to any gauge fixing. This is accomplished by providing an expression for the spatial components of the connection that allows us to isolate the nondynamical variables present among them, which can later be eliminated from the action by using their own equation of motion. As a result, we obtain the description of the phase space of general relativity in terms of manifestly $SO(n-1,1)$ [or $SO(n)$] covariant variables subject to first-class constraints only, with no second-class constraints arising during the process. Afterwards, we perform, at the covariant level, a canonical transformation to a set of variables in terms of which the above constraints take a simpler form. Finally, we impose the time gauge and make contact with the $SO(n-1)$ ADM formalism.
	\end{abstract}
	
	\maketitle
	
	\section{\label{sec:Intro} Introduction}
	The canonical analysis of general relativity has a very long history starting with attempts by Dirac himself (see for instance Refs.~\cite{RovBook,ThieBook}). However, it was not until the discovery of the ADM variables for general relativity~\cite{Arnowitt2008} that the program to canonically quantize gravity acquired a suitable and feasible form. These variables arise from the canonical analysis of the Einstein-Hilbert action through the parametrization of the spacetime metric $g_{\mu\nu}$ in terms of the lapse function $N$, the shift vector $N^a$, and the spatial metric $q_{ab}:=g_{ab}$. It turns out that in the resulting Hamiltonian form of the action both $N$ and $N^a$ play the role of Lagrange multipliers imposing the scalar (or Hamiltonian) and diffeomorphism constraints, respectively, whereas $q_{ab}$ and its canonically conjugate momentum $\tilde{p}^{ab}$--an object related to the extrinsic curvature--constitute the canonical variables that label the points of the phase space. Even though the canonical quantization program emerging from this approach has failed~\cite{Rovelli_2015}, the ADM variables have been extensively used in other instances of general relativity such as initial value problems, spacetime symmetries, asymptotic behavior of gravitational fields, numerical relativity, etc. 
	
	On the other hand, the metric formulation is not the appropriate theoretical framework to couple fermion fields to general relativity, for which we have to use the first-order formalism of the theory, where the fundamental variables are an orthonormal frame of 1-forms $e^I$ (vielbein) and an $SO(n-1,1)$ or $SO(n)$ connection 1-form $\omega^I{}_J$ depending on whether the spacetime metric has Lorentzian or Euclidean signature. The equations of motion of the theory are then obtained from the Palatini (also called Einstein-Cartan or Hilbert-Palatini) action. 
	
	The standard canonical analysis of the Palatini action involves second-class constraints, which must be either handled with the Dirac bracket~\cite{dirac1964lectures}, or explicitly solved. In 4-dimensional spacetimes, the second-class constraints are irreducible~\cite{ashtekar1991lectures} and can be explicitly solved in a manifestly $SO(3,1)$ [or $SO(4)$] covariant fashion~\cite{ashtekar1991lectures,Peldan9400}, whereas in dimensions higher than four they are reducible but can be handled using the approach of Refs.~\cite{Bodendorfer_2013,Thiemann_2013}, where the original second-class constraints are replaced with an equivalent (irreducible) set of constraints that can be explicitly solved. The second-class constraints in dimensions equal or higher than four can also be solved using the approach of Ref.~\cite{Montesinos1801}--where the second-class constraints emerging from the canonical analysis of the Holst action~\cite{Holst} are explicitly solved in a manifestly $SO(3,1)$ [or $SO(4)$] covariant fashion--because that technique is generic and {\it is not} restricted to 4-dimensional spacetimes. However, it was recently shown in Ref.~\cite{PhysRevDTBP} that it is possible to perform a manifestly $SO(3,1)$ [or $SO(4)$] covariant canonical analysis of the Holst action involving first-class constraints only, i.e., without introducing second-class constraints whatsoever in the Hamiltonian formalism. It is clear from that approach that the second-class constraints are unnecessary and superfluous for doing the canonical analysis of the Holst action, and thus they are also unnecessary for doing the Hamiltonian analysis of the 4-dimensional Palatini action as can be seen from taking the limit $\gamma\rightarrow\infty$ in Ref.~\cite{PhysRevDTBP}, where $\gamma$ is the Immirzi parameter~\cite{Immirzi9700}. 
	
	In this paper we extend the theoretical approach of Ref.~\cite{PhysRevDTBP} to higher dimensions and perform from scratch the canonical analysis of the $n$-dimensional Palatini action with a cosmological constant. In this framework, the original frame variables $e_{\mu}{}^I$ are parametrized in terms of the momentum variables, the lapse function, and the shift vector, whereas the original connection variables $\omega_{\mu}{}^I{}_J$ are expressed in terms of the configuration variables, some auxiliary fields, and some Lagrange multipliers. The outstanding aspect of this parametrization is that it straightforwardly leads to the Hamiltonian form of the $n$-dimensional Palatini action after getting rid of the auxiliary fields involved in the action. Moreover, the resulting canonical formulation is manifestly $SO(n-1,1)$ [or $SO(n)$] covariant and features first-class constraints only.
	
	This paper is organized as follows. In Sec.~\ref{sec:HA} we perform the $(n-1)+1$ decomposition of the $n$-dimensional Palatini action with or without a cosmological constant $(n\geq3)$ and provide the appropriate parametrizations of the frame and the connection. We then identify the auxiliary fields present in the action and eliminate them, thus getting the Hamiltonian form of the $n$-dimensional Palatini action with manifest local $SO(n-1,1)$ [or $SO(n)$] symmetry that involves just first-class constraints. In Sec.~\ref{Qs} we perform a canonical transformation to new $SO(n-1,1)$ [or $SO(n)$] variables that simplify the expressions of the constraints. In Sec.~\ref{TG} we impose the time gauge and obtain the $SO(n-1)$ ADM formulation of general relativity. In Sec.~\ref{Concl} we give some conclusions. In addition, in Appendix \ref{AppendixA} we discuss in detail the 3-dimensional Palatini action (for which the auxiliary fields are absent from the very beginning), and in Appendix \ref{AppendixB} we depict an alternative approach for the 4-dimensional case.
	
	\section{\label{sec:HA} Manifestly Lorentz-covariant canonical analysis}
	Let $M$ be a $n$-dimensional Lorentzian or Riemannian manifold. Points of $M$ are labeled with coordinates $x^{\alpha}$, where Greek letters $\alpha, \beta, \ldots$ represent spacetime indices. To carry out the canonical analysis, we assume that $M$ can be foliated by spacelike leaves diffeomorphic to $\Sigma$ so that $M$ is diffeomorphic to $\mathbb{R} \times \Sigma$, with $\Sigma$ being an orientable $(n-1)$-dimensional spatial manifold without boundary. We use local coordinates $(x^{\alpha})=(t,x^{a})$ adapted to this foliation of spacetime, where $t$  and $x^a$ ($a,b, \ldots=1,\ldots,n-1$) label points on $\mathbb{R}$ and $\Sigma$, respectively. In the first-order formalism, the fundamental variables are an orthonormal frame of 1-forms $e^I$ and a connection 1-form $\omega^I{}_J$ compatible with the metric $(\eta_{IJ}) := \mathrm{diag}(\sigma,1,\ldots, 1)$, $d \eta_{IJ}- \omega^K{}_I \eta_{KJ} - \omega^K{}_J \eta_{IK}=0$, and thus $\omega_{IJ}=-\omega_{JI}$ because frame indices $I,J,\ldots=0,\ldots,n-1$ are raised and lowered with $\eta_{IJ}$. For $\sigma=-1$ the frame rotation group is the Lorentz group $SO(n-1,1)$, whereas for $\sigma=+1$ it is the rotation group $SO(n)$. The weight of tensor densities is either denoted with a tilde ``$\sim$'' or explicitly mentioned somewhere in the paper. The $SO(n-1,1)$ [or $SO(n)$] totally antisymmetric tensor $\epsilon_{I_1 \cdots  I_n}$ is such that $\epsilon_{01\cdots n-1}=1$. Likewise, the totally antisymmetric spacetime tensor density of weight $+1$ ($-1$) is denoted as $\tilde{\eta}^{\alpha_{1}\cdots\alpha_{n}}$ ($\underaccent{\tilde}{\eta}_{\alpha_{1}\cdots \alpha_{n}}$) and satisfies $\tilde{\eta}^{t 1 \cdots n-1}=1$ ($\underaccent{\tilde}{\eta}_{t 1 \cdots n-1}=1$). The symmetrizer and the antisymmetrizer are defined by $V_{(\alpha \beta)}:=(V_{\alpha \beta} + V_{\beta \alpha})/2$ and $V_{[\alpha \beta]}:=(V_{\alpha \beta} - V_{\beta \alpha})/2$, respectively. ``$\wedge$'' and ``$d$'' stand for the wedge product and the exterior derivative of differential forms, correspondingly. 
		
	In the first-order formalism, general relativity with a vanishing or nonvanishing cosmological constant $\Lambda$ is described by the Palatini (or Einstein-Cartan) action\footnote{The equations of motion obtained from this action are equivalent--for nondegenerate orthonormal frames--to Einstein's equations $R_{\alpha\beta}- \frac{1}{2} R g_{\alpha\beta} + \Lambda g_{\alpha\beta}=0$.}
	\begin{equation}
			\label{Palatini}
				S[e,\omega]= \kappa \int_{M}\Bigl[ \star \left( e^{I} \wedge e^{J} \right) \wedge F_{I J} - 2\Lambda \rho \Bigr], 
	\end{equation}
	where  $F^{I}{}_{J}:=d\omega^I{}_J + \omega^{I}{}_{K} \wedge \omega^K{}_J$ is the curvature of $\omega^{I}{}_{J}$, $\rho:=(1/n!)\epsilon_{I_{1} \cdots I_{n}}e^{I_{1}}\wedge \cdots \wedge e^{I_{n}}$ is the volume form of $M$, $\kappa$ is a constant related to Newton's constant, and ``$\star$'' is the Hodge dual map given by 
	\begin{equation}
			\label{Hodge}
			\star \left( e_{I_{1}}\wedge \cdots \wedge e_{I_{k}} \right) := \frac{1}{(n-k)!}\epsilon_{I_{1}\cdots I_{k}I_{k+1}\cdots I_{n}} e^{I_{k+1}}\wedge \cdots \wedge e^{I_{n}}.
	\end{equation}
	To perform the canonical analysis of the action~\eqref{Palatini}, we first make the $(n-1)+1$ decomposition of it by expressing the frame and the connection respectively as $e^{I}=e_{t}{}^{I}dt + e_{a}{}^{I}dx^{a}$ and $\omega^{I}{}_{J}=\omega_{t}{}^{I}{}_{J}dt + \omega_{a}{}^{I}{}_{J}dx^{a}$. It is also convenient to introduce the unit normal to each leaf $\Sigma$, $n:=n_I e^I$, that fulfills $n^{I}n_{I}=\sigma$ and $n (\partial_a) =0$ (or, equivalently, $e_a{}^In_I=0$), which has the following explicit form:
	\begin{equation}
			\label{vectorn}
			n^{I}:= \frac{1}{(n-1)!\sqrt{q}}\epsilon^{I I_1 \cdots I_{n-1}}\tilde{\eta}^{t a_1 \cdots a_{n-1}} e_{a_1 I_1}  \cdots e_{a_{n-1} I_{n-1}},
	\end{equation}
	with $q:=\mathrm{det} (q_{ab}) >0$ (of weight +2), $q_{a b}:=e_{a I}e_{b}{}^{I}$ being the induced metric on $\Sigma$, whose inverse is denoted by $q^{a b}$. This object allows us to introduce the projector on the orthogonal plane to $n^{I}$ as
	\begin{equation}
			\label{proyector}
			q^{I}{}_{J}:= q^{a b}e_{a}{}^{I}e_{b J} = \delta^{I}_{J} - \sigma n^{I}n_{J}.
	\end{equation}
	
	Therefore, the $(n-1)+1$ decomposition of the action \eqref{Palatini} is given by (we recall that all spatial boundary terms will be neglected because $\Sigma$ has no boundary)
	\begin{eqnarray} 
			\label{n-1+1decomposition}
			S &=& \kappa \int_{\mathbb{R} \times \Sigma}dtd^{n-1}x \left (-2\tilde{\Pi}^{a I}n^{J}\partial_{t}\omega_{a I J} + \omega_{t I J}\tilde{\mathcal{G}}^{I J} \right. \nonumber \\
			&&+ \left. e_{t I}\tilde{\mathcal{C}}^{I} \right ),
	\end{eqnarray}
	where we have defined
	\begin{subequations}
		\begin{eqnarray}
			\label{momentum} \tilde{\Pi}^{a I} &:=& \sqrt{q}q^{a b}e_{b}{}^{I}, \\
			\label{Gauss1} \tilde{\mathcal{G}}^{I J} &:=& -2\delta^{[I}_{K} \delta^{J]}_{L} \left[ \partial_{a} \left( \tilde{\Pi}^{a K} n^{L} \right) + 2\omega_{a}{}^{K}{}_{M} \tilde{\Pi}^{a [M} n^{L]} \right]\!\!, \\
			\label{space_time_difeomorphism} \tilde{\mathcal{C}}_{I} &:=& \frac{1}{\sqrt{q}}\left[ 2\tilde{\Pi}^{a}{}_{I}\tilde{\Pi}^{b J}n^{K}F_{a b J K} + n_{I}\left( \tilde{\Pi}^{a J}\tilde{\Pi}^{b K}F_{a b J K} \right. \right. \nonumber \\
			&&- \left. 2\Lambda q \Bigr) \right]\!,
		\end{eqnarray}
	\end{subequations}
	with $F_{a b}{}^{I}{}_{J}:=\partial_{a}\omega_{b}{}^{I}{}_{J} - \partial_{b}\omega_{a}{}^{I}{}_{J} + \omega_{a}{}^{I}{}_{K}\omega_{b}{}^{K}{}_{J} - \omega_{b}{}^{I}{}_{K}\omega_{a}{}^{K}{}_{J} $ being the curvature of $\omega_{a}{}^{I}{}_{J}$ and where we have also suppressed a wedge product between $dt$ and $d^{n-1}x:=dx^1 \wedge \cdots \wedge dx^{n-1}$ in~\eqref{n-1+1decomposition} to simplify notation. 
	
	To continue our analysis, we express $e_t{}^I$ in terms of the lapse function $N$ and the shift vector $N^{a}$~\cite{Arnowitt2008} as
	\begin{equation}
			\label{temporalpart}
			e_{t}{}^{I} = Nn^{I} + N^{a}e_{a}{}^{I},
	\end{equation}
	and compute the inverse of the expression~\eqref{momentum}
	\begin{equation}
			\label{spatial_n-bein_in_terms_of_Pi}
			e_{a}{}^{I}=h^{\frac{1}{2(n-2)}} \uac{\uac{h}}_{a b}\tilde{\Pi}^{b I},
	\end{equation}
where $\uac{\uac{h}}_{a b}$ is the inverse of $\tilde{\tilde{h}}^{ab}:=\tilde{\Pi}^{a I}\tilde{\Pi}^{b}{}_{I}$ and $h:=\mathrm{det}(\tilde{\tilde{h}}^{a b})$ has weight $2(n-2)$. Notice that the right-hand side of~\eqref{spatial_n-bein_in_terms_of_Pi} is a function of $\tilde{\Pi}^{aI}$ only. As a consequence of this, $n^{I}$ in~\eqref{vectorn} can also be expressed in terms of $\tilde{\Pi}^{a I}$ as
	\begin{equation}
			\label{vectorn_in_terms_of_Pi}
			n_{I}=\frac{1}{(n-1)! \sqrt{h}}\epsilon_{II_1 \cdots I_{n-1}}\underaccent{\tilde}{\eta}_{t a_1 \cdots a_{n-1}} \tilde{\Pi}^{a_1 I_1}  \cdots \tilde{\Pi}^{a_{n-1} I_{n-1}}.
	\end{equation}
Substituting~\eqref{spatial_n-bein_in_terms_of_Pi} and~\eqref{vectorn_in_terms_of_Pi} into the right-hand side of~\eqref{temporalpart} we can reinterpret $e_t{}^I$ as a function of the $n^2$ variables $N$, $N^a$, and $\tilde{\Pi}^{a I}$. With this in mind, relations~\eqref{temporalpart} and~\eqref{spatial_n-bein_in_terms_of_Pi} define a one-to-one map from the $n^2$ variables $N$, $N^a$, and $\tilde{\Pi}^{a I}$ to the original $n^2$ frame components $e_{\alpha}{}^I$. The inverse map that sends $e_{\alpha}{}^I$ to $N$, $N^a$, and $\tilde{\Pi}^{a I}$ is given by~\eqref{momentum} together with
	\begin{subequations}
			\begin{eqnarray}
		N&=& \sigma e_t{}^I n_I, \\
		N^a&=& q^{ab}e_t{}^I e_{bI},
		\end{eqnarray}
	\end{subequations}
where $n_I$ must be understood as that given by~\eqref{vectorn}.
	
Therefore, using~\eqref{temporalpart},~\eqref{spatial_n-bein_in_terms_of_Pi}, and~\eqref{vectorn_in_terms_of_Pi}, the action~\eqref{n-1+1decomposition} acquires the form\footnote{From~\eqref{spatial_n-bein_in_terms_of_Pi} we get $h=q^{n-2}$, and thus $\sqrt{q}=h^{\frac{1}{2 (n-2)}}$.}
	\begin{eqnarray}\label{action_in_terms_of_Pi}
		S &=&\kappa \int_{\mathbb{R} \times \Sigma}dt d^{n-1}x \left ( -2\tilde{\Pi}^{a I}n^{J}\partial_{t}\omega_{a I J} + \omega_{t I J}\tilde{\mathcal{G}}^{I J} \right. \nonumber \\ 
		&& - \left. N^{a}\tilde{\mathcal{V}}_{a} - \underaccent{\tilde}{N} \tilde{\tilde{\mathcal{C}}} \right ),
	\end{eqnarray}
with 
	\begin{subequations}
	\begin{eqnarray}
	\label{vector_constraint} \tilde{\mathcal{V}}_{a}&:=& -2\tilde{\Pi}^{bI}n^{J} F_{a b I J}, \\
	\label{scalar_constraint} \tilde{\tilde{\mathcal{C}}} &:=& -\sigma \tilde{\Pi}^{a I}\tilde{\Pi}^{b J}F_{a b I J} + 2\sigma h^{\frac{1}{n-2}}\Lambda, \\
	\underaccent{\tilde}{N}&:=& h^{-\frac{1}{2(n-2)}} N.
	\end{eqnarray}
	\end{subequations}
	
	For future purposes, we introduce the covariant derivative $\nabla_{a}$ defined on each leaf $\Sigma$ that annihilates $e_{a}{}^{I}$ through
	\begin{equation}
			\label{covariant_derivative}
			\nabla_{a}e_{b}{}^{I}:= \partial_{a}e_{b}{}^{I} + \Gamma_{a}{}^{I}{}_{J} e_{b}{}^{J}  - \Gamma^{c}{}_{ab} e_{c}{}^{I} = 0, 
	\end{equation}
	with $\Gamma_{aIJ}=-\Gamma_{aJI}$ and $\Gamma^{a}{}_{bc}=\Gamma^{a}{}_{cb}$. These are $n(n-1)^{2}$ inhomogeneous linear equations for $n(n-1)^{2}/2$ unknowns $\Gamma_{aIJ}$ and $n(n-1)^{2}/2$ unknowns $\Gamma^a{}_{bc}$, so that the solution is unique. It turns out that $\Gamma^{a}{}_{b c}$ are the Christoffel symbols associated with the induced metric $q_{ab}$ on $\Sigma$, whereas the explicit solution for $\Gamma_{aIJ}$ is given by
	\begin{eqnarray}
			\label{solution_of_Gamma}
			 \Gamma_{a I J} &=& q^{b c}e_{b [I|} \left( \partial_{a}e_{c|J]} - \partial_{c}e_{a|J]} \right) + \sigma q^{b c}e_{b [I}n_{J]}n_{K} \nonumber \\
			&& \times \left( \partial_{a} e_{c}{}^{K} + \partial_{c} e_{a}{}^{K} \right) + q^{b c} q^{d f}e_{a K}e_{b [I}e_{|d|J]}\partial_{f}e_{c}{}^{K}. \notag \\
	\end{eqnarray}
	Furthermore, from~\eqref{momentum} and \eqref{covariant_derivative}, we find that the operator $\nabla_{a}$ annihilates $\tilde{\Pi}^{a I}$ as well 
	\begin{eqnarray}
			\label{covariant_derivative_Pi}
			\nabla_{a}\tilde{\Pi}^{b I}&=& \partial_{a}\tilde{\Pi}^{b I} + \Gamma_{a}{}^{I}{}_{J}\tilde{\Pi}^{b J} + \Gamma^{b}{}_{a c}\tilde{\Pi}^{c I} - \Gamma^{c}{}_{a c}\tilde{\Pi}^{b I} =0. \nonumber\\
	\end{eqnarray}
	Either by solving this equation similarly as we did for~\eqref{covariant_derivative} or simply by substituting~\eqref{spatial_n-bein_in_terms_of_Pi} into the right-hand-side of~\eqref{solution_of_Gamma}, we find
	\begin{eqnarray}
			\label{solution_of_Gamma_in_terms_of_Pi}
			\Gamma_{a I J} &=&  \uac{\uac{h}}_{a b}\tilde{\Pi}^{c}{}_{[I}\partial_{|c|}\tilde{\Pi}^{b}{}_{J]} +  \uac{\uac{h}}_{a b} \uac{\uac{h}}_{c d}\tilde{\Pi}^{c}{}_{K}\tilde{\Pi}^{b}{}_{[I}\tilde{\Pi}^{f}{}_{J]}\partial_{f}\tilde{\Pi}^{d K} \nonumber \\
			&& + \uac{\uac{h}}_{b c}\tilde{\Pi}^{b}{}_{[I}\partial_{|a|}\tilde{\Pi}^{c}{}_{J]} - \uac{\uac{h}}_{a b}\uac{\uac{h}}_{c d}\tilde{\Pi}^{b}{}_{K}\tilde{\Pi}^{c}{}_{[I}\tilde{\Pi}^{f}{}_{J]}\partial_{f}\tilde{\Pi}^{d K} \nonumber \\
			&& - \sigma \uac{\uac{h}}_{a b}\tilde{\Pi}^{c}{}_{[I}n_{J]}n_{K}\partial_{c}\tilde{\Pi}^{b K} + \sigma \uac{\uac{h}}_{b c}\tilde{\Pi}^{b}{}_{[I}n_{J]}n_{K}\partial_{a}\tilde{\Pi}^{c K}. \notag\\
	\end{eqnarray}

	Now, following the same approach of Refs.~\cite{Montesinos1801,PhysRevDTBP}, we realize that the term involving $\partial_t \omega_{aIJ}$ in~\eqref{action_in_terms_of_Pi} can be written as
	\begin{equation}
			\label{kinetic_term}
			-2\tilde{\Pi}^{a I}n^{J}\partial_{t}\omega_{a I J} = 2 \tilde{\Pi}^{a I}\partial_{t} \left ( W_{a}{}^{b}{}_{I J K}\omega_{b}{}^{J K} \right ),
	\end{equation}
with $W_{a}{}^{b}{}_{IJK} = -W_{a}{}^{b}{}_{IKJ}$ given by
\begin{equation}
			\label{equation_for_W}
			W_{a}{}^{b}{}_{IJK}:= -\left( \delta_{a}^{b} \eta_{I [J}n_{K]} + n_{I} \uac{\uac{h}}_{a c}\tilde{\Pi}^{c}{}_{[J}\tilde{\Pi}^{b}{}_{K]} \right).
	\end{equation}
It is worthwhile to remark that the equality~\eqref{kinetic_term} {\it is exact}. That is to say, neither temporal nor spatial boundary terms have been neglected. The relation~\eqref{kinetic_term} clearly suggests to define the $n(n-1)$ configuration variables 
           \begin{equation}\label{q_equation}
			 {\cal{Q}}_{a I}:=W_{a}{}^{b}{}_{I J K}\omega_{b}{}^{J K},
			 \end{equation}
which thus are canonically conjugate to $\tilde{\Pi}^{aI}$. The variables ${\cal{Q}}_{aI}$ embody the combination of the components of the connection $\omega_{a}{}^{IJ}$ contributing to the dynamical variables of the theory; those variables are precisely singled out by the object $W_{a}{}^{b}{}_{I J K}$. We can interpret~\eqref{q_equation} as $n(n-1)$ linear equations for $n(n-1)^{2}/2$ unknowns $\omega_{aIJ}$. In consequence, the solution for $\omega_{aIJ}$ must involve $n(n-1)^{2}/2-n(n-1)=n(n-1)(n-3)/2$ free variables. Let us call these variables $\uac{\uac{\lambda}}_{abc}$, which satisfy $\uac{\uac{\lambda}}_{abc}= - \uac{\uac{\lambda}}_{acb}$ and the traceless condition $\uac{\uac{\lambda}}_{abc} \tilde{\tilde{h}}^{ab}=0$; both conditions guarantee the right amount of independent variables that $\uac{\uac{\lambda}}_{abc}$ must contain. The solution for $\omega_{aIJ}$ can be expressed as
	\begin{equation}
			\label{solution_omega}
			\omega_{aIJ} = M_{a}{}^{b}{}_{IJK} {\cal{Q}}_{b}{}^{K} +  \tilde{\tilde{N}}_a{}^{bcd}{}_{IJ}\uac{\uac{\lambda}}_{bcd},
	\end{equation}
	with
	\begin{eqnarray}
			 M_{a}{}^{b}{}_{I J K} &:=& \frac{2 \sigma}{(n-2)}\Big{[} (n-2)\delta^{b}_{a}n_{[I}\eta_{J] K} +  \uac{\uac{h}}_{a c}\tilde{\Pi}^{c}{}_{[I}\tilde{\Pi}^{b}{}_{J]}n_{K} \Big{]},\notag\\
			 \label{equation_M}\\
			 \tilde{\tilde{N}}_a{}^{bcd}{}_{IJ}&:=&\left(\delta_a^b\delta_e^{[c}\delta_f^{d]}-\frac{2}{n-2}\uac{\uac{h}}_{ae}\tilde{\tilde{h}}^{b[c}\delta_f^{d]}\right)\tilde{\Pi}^{e}{}_{[I}\tilde{\Pi}^{f}{}_{J]}. \label{equation_N}
	\end{eqnarray}
Notice that $M_{a}{}^{b}{}_{IJK}$ and $\tilde{\tilde{N}}_a{}^{bcd}{}_{IJ}$ satisfy $M_{a}{}^{b}{}_{IJK} = -M_{a}{}^{b}{}_{JIK}$, $\tilde{\tilde{N}}_a{}^{bcd}{}_{IJ}=-\tilde{\tilde{N}}_a{}^{bcd}{}_{JI}=-\tilde{\tilde{N}}_a{}^{bdc}{}_{IJ}$, and $\uac{\uac{h}}_{b c}\tilde{\tilde{N}}_a{}^{bcd}{}_{IJ}=0$.
We point out that the variables $\uac{\uac{\lambda}}_{abc}$ are present in~\eqref{solution_omega} only for $n \geq 4$. When $n=3$, there are no variables $\uac{\uac{\lambda}}_{abc}$ in~\eqref{solution_omega} because in that case both the number of equations contained in the expression~\eqref{q_equation} and the number of unknowns $\omega_{aIJ}$ are equal to six. Despite the fact that there are no variables $\uac{\uac{\lambda}}_{abc}$ for $n=3$, we will show in Appendix~\ref{AppendixA} that the final canonical analysis for $n=3$ has exactly the same form as the case $n \geq 4$. Let us consider $n >3$ from now on in this section. For the sake of completeness, we define the tensor density $\uac{\uac{U}}_{abc}{}^{dIJ}$ with the properties $\uac{\uac{U}}_{abc}{}^{dIJ} = - \uac{\uac{U}}_{acb}{}^{dIJ}= - \uac{\uac{U}}_{abc}{}^{dJI}$ and $\tilde{\tilde{h}}^{ab} \uac{\uac{U}}_{abc}{}^{dIJ}=0$ as follows:
	\begin{equation}
			\label{equation_for_U}
			\uac{\uac{U}}_{abc}{}^{dIJ}:= \left( \delta^{d}_{a}\uac{\uac{h}}_{e[b}\uac{\uac{h}}_{c]f} - \frac{2}{n-2}\uac{\uac{h}}_{a [b}\uac{\uac{h}}_{c] f}\delta^{d}_{e} \right)\tilde{\Pi}^{e [I}\tilde{\Pi}^{|f|J]}.
	\end{equation}
It is related to $\tilde{\tilde{N}}_a{}^{bcd}{}_{IJ}$ by
\begin{equation}
	\tilde{\tilde{h}}^{ea}\tilde{\tilde{h}}^{gb}\tilde{\tilde{h}}^{hc}\uac{\uac{h}}_{fd}\uac{\uac{U}}_{abc}{}^{dIJ}=\tilde{\tilde{N}}_f{}^{eghIJ}.
\end{equation}
The objects \eqref{equation_for_W}, \eqref{equation_M}, \eqref{equation_N} and \eqref{equation_for_U} all together fulfill the orthogonality relations 
	\begin{subequations}
	\begin{eqnarray} 
	W_{a}{}^{c I KL}M_{c}{}^{b}{}_{KLJ} &=& \delta^{b}_{a}\delta^{I}_{J},\label{WM} \\
	 \label{orthogonal_relation_NU} \uac{\uac{U}}_{cde}{}^{gIJ}  \tilde{\tilde{N}}_g{}^{fab}{}_{IJ} &=&\delta^f_c \delta^{[a}_{d}\delta^{b]}_{e}\nonumber\\
	&& - \frac{1}{n-2} \left ( \uac{\uac{h}}_{cd} \tilde{\tilde{h}}^{f[a} \delta^{b]}_e - \uac{\uac{h}}_{ce} \tilde{\tilde{h}}^{f[a} \delta^{b]}_d \right ), \notag \\
	&& \\
	 \label{orthogonal_relation_WN} W_{a}{}^{f}{}_{IJK}\tilde{\tilde{N}}_f{}^{bcdJK} &=& 0, \\
	 \uac{\uac{U}}_{abc}{}^{dIJ}M_{d}{}^{e}{}_{IJK} &=& 0.
	\end{eqnarray}
	\end{subequations}
The presence of the second term on the right-hand side of~\eqref{orthogonal_relation_NU} is a consequence of both traceless conditions $\uac{\uac{h}}_{b c}\tilde{\tilde{N}}_a{}^{bcd}{}_{IJ}=0$ and $\tilde{\tilde{h}}^{ab} \uac{\uac{U}}_{abc}{}^{dIJ}=0$. Using~\eqref{solution_omega} together with the relations \eqref{WM} and \eqref{orthogonal_relation_NU}, we get~\eqref{q_equation} as well as 
\begin{eqnarray}\label{aux_fields}
\uac{\uac{\lambda}}_{abc} = \uac{\uac{U}}_{abc}{}^{dIJ} \omega_{dIJ},
\end{eqnarray}
which shows that ${\cal{Q}}_{aI}$ and $\uac{\uac{\lambda}}_{abc}$ are independent variables among themselves. Furthermore, we have the completeness relation
\begin{eqnarray}
	M_a{}^c{}_{IJM} W_c{}^{bMKL} + \tilde{\tilde{N}}_a{}^{cdf}{}_{IJ}\uac{\uac{U}}_{cdf}{}^{bKL}  &=& \delta^b_a \delta^K_{[I} \delta^L_{J]}.
\end{eqnarray}
   
	Now, we replace $\omega_a{}^I{}_J$ with ${\cal{Q}}_{aI}$ and $\uac{\uac{\lambda}}_{abc}$ by substituting~\eqref{solution_omega}  into the action principle~\eqref{action_in_terms_of_Pi}  and obtain 
	\begin{eqnarray}\label{action_in_terms_of_lambda}
	S &=&\kappa \int_{\mathbb{R} \times \Sigma}dt d^{n-1}x \left ( 2\tilde{\Pi}^{a I}\partial_{t} {\cal{Q}}_{a I} + \omega_{t I J}\tilde{\mathcal{G}}^{I J} \right. \nonumber \\ 
	&& \left. - N^{a}\tilde{\mathcal{V}}_{a} - \underaccent{\tilde}{N} \tilde{\tilde{\mathcal{C}}} \right ),
	\end{eqnarray} 
	with
	\begin{widetext}
	\begin{subequations}
	\begin{eqnarray}
			\label{Gauss2} \tilde{\mathcal{G}}^{I J} &=& 2 \tilde{\Pi}^{a [I} {\cal{Q}}_{a}{}^{J]} + 4 \delta^{I}_{[K}\delta^{J}_{L]} \tilde{\Pi}^{a [K} n^{M]} \Gamma_{a}{}^{L}{}_{M}, \\
			\label{vector2} \tilde{\mathcal{V}}_{a} &=& 2 \left( 2 \tilde{\Pi}^{b I}\partial_{[a} {\cal{Q}}_{b] I} - {\cal{Q}}_{a I} \partial_{b}\tilde{\Pi}^{b I} \right) + \tilde{\mathcal{G}}_{I J}\left( M_{a}{}^{bIJK} {\cal{Q}}_{bK} + \tilde{\tilde{N}}_a{}^{bcdIJ} \uac{\uac{\lambda}}_{bcd} \right),\\
			\label{scalar2}  \tilde{\tilde{\mathcal{C}}} &=& - \sigma \tilde{\Pi}^{a I} \tilde{\Pi}^{b J} R_{a b I J} + 2 \tilde{\Pi}^{a [I}\tilde{\Pi}^{|b|J]} \left[ {\cal{Q}}_{aI} {\cal{Q}}_{bJ} + 2 {\cal{Q}}_{aI} \Gamma_{bJK} n^{K} + \Gamma_{a I L}\Gamma_{b J K}n^{K}n^{L} \right] + 2\sigma \Lambda h^{\frac{1}{(n-2)}} + 2 \tilde{\Pi}^{a I} n^{J} \nabla_{a}\tilde{\mathcal{G}}_{IJ} \nonumber \\
			&& - \frac{(n-3)}{(n-2)} \sigma n^{I}\tilde{\mathcal{G}}^{J}{}_{K} n^{K}\tilde{\mathcal{G}}_{I J} + \sigma \tilde{\tilde{h}}^{db} \tilde{\tilde{h}}^{cf} \tilde{\tilde{h}}^{e a} \left( \uac{\uac{\lambda}}_{abc} - \uac{\uac{U}}_{abc}{}^{h}{}_{KL}\Gamma_{h}{}^{KL} \right) \left( \uac{\uac{\lambda}}_{dfe} - \uac{\uac{U}}_{dfe}{}^{g}{}_{IJ}\Gamma_{g}{}^{IJ} \right),
	\end{eqnarray}
	\end{subequations}
	\end{widetext}
	where $R_{ab}{}^I{}_J:= \partial_{a}\Gamma_b{}^I{}_J - \partial_{b}\Gamma_a{}^I{}_J + \Gamma_a{}^I{}_K \Gamma_b{}^K{}_J - \Gamma_b{}^I{}_K \Gamma_a{}^K{}_J$ is the curvature of the connection $\Gamma_a{}^I{}_J$.
	
	It is remarkable that $\tilde{\mathcal{G}}^{I J}$--given by~\eqref{Gauss2}--involves no $\uac{\uac{\lambda}}_{abc}$. It is also surprising that $\tilde{\mathcal{V}}_{a}$ and 
	$\tilde{\tilde{\mathcal{C}}}$--given correspondingly by~\eqref{vector2} and~\eqref{scalar2}--contain no spatial derivatives of $\uac{\uac{\lambda}}_{abc}$, because~\eqref{vector_constraint} and~\eqref{scalar_constraint} contain spatial derivatives of $\omega_a{}^I{}_J$. By inspection, it is pretty obvious that the variables $\uac{\uac{\lambda}}_{abc}$ are auxiliary fields~\cite{HennBook}. At this point, there are two, equivalent, ways to continue. The first way consists in to first fix the variables $\uac{\uac{\lambda}}_{abc}$ by using their equation of motion and then to substitute them back into the action~\eqref{action_in_terms_of_lambda}. Next, a redefinition of the Lagrange multiplier in front of the Gauss constraint $\tilde{\mathcal{G}}^{I J}$ is required (this way was followed in Ref~\cite{PhysRevDTBP}). The second way consists in first to redefine the Lagrange multiplier in front of $\tilde{\mathcal{G}}^{I J}$ and then to get rid of the auxiliary fields $\uac{\uac{\lambda}}_{abc}$. We will follow the second way. Then, factoring out all terms in $\tilde{\mathcal{V}}_{a}$ and 
	$\tilde{\tilde{\mathcal{C}}}$ involving $\tilde{\mathcal{G}}^{I J}$, we get
	\begin{eqnarray}\label{action_with_Lambda}
		S &=&\kappa \int_{\mathbb{R} \times \Sigma}dt d^{n-1}x\left ( 2\tilde{\Pi}^{a I}\partial_{t} {\cal{Q}}_{a I} -\Lambda_{IJ} \tilde{\mathcal{G}}^{I J} \right. \nonumber \\ 
		&& \left. -2N^{a}\tilde{\mathcal{D}}_{a} - \underaccent{\tilde}{N} \tilde{\tilde{\mathcal{S}}} \right ),
	\end{eqnarray}
	with
	\begin{subequations}
	\begin{eqnarray}
	\label{Gauss3} \tilde{\mathcal{G}}^{I J} &=& 2 \tilde{\Pi}^{a [I} {\cal{Q}}_{a}{}^{J]} + 4 \delta^{I}_{[K}\delta^{J}_{L]} \tilde{\Pi}^{a [K} n^{M]} \Gamma_{a}{}^{L}{}_{M}, \\
	\label{diffeomorphism} \tilde{\mathcal{D}}_{a} &:=& 2\tilde{\Pi}^{b I} \partial_{[a} {\cal{Q}}_{b] I} - {\cal{Q}}_{a}{}^{I}\partial_{b}\tilde{\Pi}^{b}{}_{I}, \\ 
	\label{scalar4} \tilde{\tilde{\mathcal{S}}} &:=& -\sigma \tilde{\Pi}^{a I}\tilde{\Pi}^{b J}R_{a b I J} + 2\tilde{\Pi}^{a [I}\tilde{\Pi}^{|b|J]} \left ( {\cal{Q}}_{aI} {\cal{Q}}_{bJ} \right. \nonumber \\
	&&  \left. +2 {\cal{Q}}_{aI} \Gamma_{bJK}n^{K} + \Gamma_{a I K}\Gamma_{bJL} n^{K}n^{L} \right ) + 2\sigma h^{\frac{1}{n-2}} \Lambda \nonumber\\
	&& + \sigma \tilde{\tilde{h}}^{db} \tilde{\tilde{h}}^{cf} \tilde{\tilde{h}}^{ea} \left( \uac{\uac{ \lambda}}_{abc} - \uac{\uac{U}}_{abc}{}^{h}{}_{KL}\Gamma_{h}{}^{K L} \right) \nonumber\\
	&& \times \left( \uac{\uac{\lambda}}_{dfe} - \uac{\uac{U}}_{dfe}{}^{g}{}_{I J}\Gamma_{g}{}^{IJ} \right),	
	\end{eqnarray}
	\end{subequations}
where $\tilde{\mathcal{D}}_{a}$ and $\tilde{\tilde{\mathcal{S}}}$ are the diffeomorphism and Hamiltonian constraints, respectively. Also, as promised, we have replaced $\omega_{tIJ}$ with $\Lambda_{IJ}$ via the field redefinition
\begin{eqnarray}\label{Lambda}
	\omega_{tIJ} &=& -\Lambda_{IJ} + N^a \left( M_a{}^b{}_ {IJK} {\cal{Q}}_b{}^K +   \tilde{\tilde{N}}_a{}^{bcd}{}_{IJ}\uac{\uac{\lambda}}_{bcd} \right) \nonumber\\
	&& - 2  \tilde{\Pi}^{a}{}_{[I} n_{J]} \nabla_a \underaccent{\tilde}{N} - \sigma\frac{(n-3)}{(n-2)} \underaccent{\tilde}{N} n_{[I} \tilde{\mathcal{G}}_{J]K} n^K.
\end{eqnarray}
Therefore, the original connection variables $\omega_{\alpha}{}^{IJ}$ have been replaced with the independent variables ${\cal{Q}}_a{}^I$, $\uac{\uac{\lambda}}_{abc}$ (satisfying the properties already mentioned for them), and $\Lambda_{IJ}$. It is clear by now that $\uac{\uac{\lambda}}_{abc}$ are auxiliary fields that can be eliminated by using their own equation of motion. In fact, by making the variation of the action~\eqref{action_with_Lambda} with respect to $\uac{\uac{\lambda}}_{abc}$ (taking into account the properties for them), we have 
\begin{eqnarray}
			\label{variation_of_action_respect_Lambda}
			\underaccent{\tilde}{N} \tilde{\tilde{h}}^{d [b}\tilde{\tilde{h}}^{c] e}\tilde{\tilde{h}}^{a f}\left( \uac{\uac{\lambda}}_{dfe} - \uac{\uac{U}}_{dfe}{}^{g}{}_{I J}\Gamma_{g}{}^{I J} \right) =0,
	\end{eqnarray}
which implies
\begin{eqnarray}
			\label{solution_for_Lambda}
			&& \uac{\uac{\lambda}}_{abc} = \uac{\uac{U}}_{abc}{}^{d}{}_{I J}\Gamma_{d}{}^{IJ}.	
			\end{eqnarray}
Substituting back $\uac{\uac{\lambda}}_{abc}$ into~\eqref{action_with_Lambda}, we arrive at the Hamiltonian form of the $n$-dimensional Palatini action with a cosmological constant $\Lambda$:
\begin{eqnarray}\label{final_action}
			 S &=&\kappa \int_{\mathbb{R} \times \Sigma}dt d^{n-1}x\left ( 2\tilde{\Pi}^{a I}\partial_{t} {\cal{Q}}_{aI} -\Lambda_{IJ} \tilde{\mathcal{G}}^{I J} \right. \nonumber \\ 
			&& \left. -2N^{a}\tilde{\mathcal{D}}_{a} - \underaccent{\tilde}{N} \tilde{\tilde{\mathcal{H}}} \right ),
	\end{eqnarray}
	with the Gauss, diffeomorphism and scalar constraints given by
	\begin{subequations}
		\begin{eqnarray}
			\label{Gauss4} \tilde{\mathcal{G}}^{IJ} &=&  2 \tilde{\Pi}^{a [I} {\cal{Q}}_{a}{}^{J]} + 4 \delta^{I}_{[K}\delta^{J}_{L]} \tilde{\Pi}^{a [K} n^{M]} \Gamma_{a}{}^{L}{}_{M}, \\
			\label{diffeomorphism2} \tilde{\mathcal{D}}_{a} &=& 2\tilde{\Pi}^{b I} \partial_{[a} {\cal{Q}}_{b] I} - {\cal{Q}}_{a}{}^{I}\partial_{b}\tilde{\Pi}^{b}{}_{I}, \\ 
			\label{scalar3} \tilde{\tilde{\mathcal{H}}} &:=& - \sigma \tilde{\Pi}^{a I}\tilde{\Pi}^{b J}R_{a b I J} +  2 \tilde{\Pi}^{a [I}\tilde{\Pi}^{|b|J]} \left ( {\cal{Q}}_{aI} {\cal{Q}}_{bJ} 
			\right. \nonumber \\
			&& \left. +2 {\cal{Q}}_{aI} \Gamma_{bJK} n^{K} + \Gamma_{aIK} \Gamma_{bJL} n^{K} n^{L} \right ) + 2 \sigma h^{\frac{1}{n-2}} \Lambda, \notag \\
		\end{eqnarray}
	\end{subequations}
	respectively. It is worth mentioning that, although the spacetime dimension $n$ shows up in the term involving the $(n-2)$-th root of $h$ in \eqref{scalar3}, the constraints \eqref{Gauss4}-\eqref{scalar3} take exactly the same form in all spacetime dimensions. For $\Lambda=0$, the form of the constraints is actually independent of the spacetime dimension.
	
	Therefore, we have obtained a manifestly Lorentz-covariant Hamiltonian formulation~\eqref{final_action} for the Palatini action~\eqref{Palatini}. This Hamiltonian form of the action emerged from parametrizing the original frame variables $e_{\alpha}{}^I$ in terms of the momentum variables $\tilde{\Pi}^{aI}$, the lapse $N$, and the shift $N^a$ as given by~\eqref{temporalpart}--\eqref{spatial_n-bein_in_terms_of_Pi}, whereas the original connection variables $\omega_{\alpha}{}^I{}_J$ have been parametrized in terms of the configuration variables ${\cal{Q}}_a{}^I$, the auxiliary fields $\uac{\uac{\lambda}}_{abc}$, and the Lagrange multipliers $\Lambda_{IJ}$ as depicted in~\eqref{solution_omega} and~\eqref{Lambda}. 
	
	Notice that the map from $\omega_a{}^I{}_J$ to ${\cal{Q}}_a{}^I$ and $\uac{\uac{\lambda}}_{abc}$ through~\eqref{q_equation} and~\eqref{aux_fields}, with inverse map given by~\eqref{solution_omega}, can be seen as a change of variables. Nevertheless, as is clear from~\eqref{kinetic_term} and~\eqref{q_equation}, the presymplectic structure present in~\eqref{action_in_terms_of_Pi} becomes the canonical symplectic structure present in~\eqref{action_in_terms_of_lambda} when such a map is used. Therefore, we reach a smaller phase-space and simultaneously parametrize it with manifestly Lorentz-covariant canonical variables (${\cal{Q}}_a{}^I$, $\tilde{\Pi}^{a}{}_I$). The reduction map is given by $(\omega_a{}^I{}_J,  \tilde{\Pi}^{a}{}_I) \longmapsto ({\cal{Q}}_a{}^I, \tilde{\Pi}^{a}{}_I)$ using~\eqref{q_equation}. This reduction process leaves the null directions of the presymplectic structure~\eqref{action_in_terms_of_Pi} out of the canonical symplectic structure present in~\eqref{action_in_terms_of_lambda}. The null directions are clearly along $\uac{\uac{\lambda}}_{abc}$, which turn out to be auxiliary fields that can be eliminated from the action by using their own equation of motion. The variables $\Lambda_{IJ}$, $N^a$, and $\uac{N}$ are Lagrange multipliers imposing the $SO(n-1,1)$ [or $SO(n)$] Gauss, diffeomorphism, and scalar constraints; respectively. These constraints depend on the phase space variables $({\cal{Q}}_a{}^I, {\tilde \Pi}^a{}_I)$ satisfying the Poisson brackets
	\begin{eqnarray}
	\{ {\cal{Q}}_a{}^I (t,x) , {\tilde \Pi}^b{}_J (t,y) \} = \frac{1}{2 \kappa} \delta^b_a \delta^I_J \delta^{n-1} (x,y).
	\end{eqnarray}
We close this section with two remarks:  
\begin{itemize}
	\item[(i)] For $4$-dimensional spacetimes, the canonical description of general relativity with a cosmological constant given in~\eqref{final_action} is the same as the one obtained from the canonical variables for the Holst action through a canonical transformation (see Sec. IV of Ref.~\cite{PhysRevDTBP}). 
	\item[(ii)] As shown in Appendix~\ref{AppendixA}, for $3$-dimensional spacetimes there are no auxiliary fields $\uac{\uac{\lambda}}_{abc}$ (notice that $\uac{\uac{U}}_{abc}{}^{dIJ}$ identically vanishes for $n=3$, as for any object with the same symmetries of $\uac{\uac{\lambda}}_{abc}$ in three of its spatial indices). In spite of this, the resulting Hamiltonian form of the theory has exactly the same structure given by~\eqref{final_action}.
\end{itemize}

\section{Other manifestly Lorentz-covariant phase-space variables}\label{Qs}
It is important to emphasize that the manifestly Lorentz-covariant canonical analysis of general relativity with a cosmological constant embodied in the action~\eqref{final_action} {\it is not} the canonical description of the Palatini action given in Refs.~\cite{Bodendorfer_2013,Thiemann_2013}. We show in what follows that the latter can be obtained from our Hamiltonian formulation through a very simple canonical transformation leaving the momentum  $\tilde{\Pi}^{a I}$ unchanged: $({\cal{Q}}_{a I}, \tilde{\Pi}^{a I}) \longmapsto (Q_{a I}, \tilde{\Pi}^{a I})$. Both configuration variables are related to each other by 
\begin{equation}
			\label{canonical_transformation}
			Q_{aI}= {\cal{Q}}_{aI} - W_{a}{}^{b}{}_{IJK}\Gamma_{b}{}^{J K}.
	\end{equation}
	This transformation is indeed canonical because
	\begin{eqnarray}\label{prove_canonical_transformation}
2 \tilde{\Pi}^{a I}\partial_{t}Q_{aI} &=& 2 \tilde{\Pi}^{a I}\partial_{t} {\cal{Q}}_{aI} + \partial_{a} \left ( 2 n_{I}\partial_{t}\tilde{\Pi}^{a I} \right ),
\end{eqnarray}
and since $\Sigma$ has no boundary, the last term of the equality~\eqref{prove_canonical_transformation} does not contribute to the Hamiltonian action. More precisely, using~\eqref{canonical_transformation}, the action~\eqref{final_action} acquires the form
\begin{eqnarray}\label{Thiemann_action}
			S &= & \kappa \int_{\mathbb{R} \times \Sigma}dt d^{n-1}x\left ( 2\tilde{\Pi}^{a I}\partial_{t}Q_{a I} -\Lambda_{IJ} \tilde{\mathcal{G}}^{I J} \right. \nonumber \\ 
			&& \left. - 2 N^{a}\tilde{\mathcal{D}}_{a} - \underaccent{\tilde}{N} \tilde{\tilde{\mathcal{H}}} \right ),
	\end{eqnarray}
	with
		\begin{subequations}
					\begin{eqnarray}
			\label{GaussQ} \tilde{\mathcal{G}}^{I J} &=& 2 \tilde{\Pi}^{a [I}Q_{a}{}^{J]}, \\
			\label{diffeomorphismQ} \tilde{\mathcal{D}}_{a} &=& 2\tilde{\Pi}^{b I} \partial_{[a} Q_{b] I} - Q_{a}{}^{I}\partial_{b}\tilde{\Pi}^{b}{}_{I}, \\
			\label{scalarQ} \tilde{\tilde{\mathcal{H}}} &=& - \sigma \tilde{\Pi}^{a I} \tilde{\Pi}^{b J}R_{a b I J} +  2 \tilde{\Pi}^{a [I}\tilde{\Pi}^{|b|J]}Q_{a I}Q_{b J} \nonumber \\
			&& + 2 \sigma h^{\frac{1}{(n-2)}}\Lambda.
			\end{eqnarray}
		\end{subequations}
	This is the formulation obtained in Ref.~\cite{Bodendorfer_2013,Thiemann_2013} through a lengthy process of solving the second-class constraints involved there. Notice also that the canonical variables $(Q_{a I}, \tilde{\Pi}^{a I})$ are $SO(n-1,1)$ [or $SO(n)$] vectors. 
	
	Alternatively, the manifestly Lorentz-covariant Hamiltonian formulation~\eqref{Thiemann_action} can also be directly obtained from~\eqref{action_in_terms_of_Pi} by following an analogous procedure to that developed in Sec.~\ref{sec:HA}. To achieve this, we have to handle the equality~\eqref{kinetic_term} as follows:
	\begin{eqnarray}
			\label{kinetic_term_Q}
			- 2 \tilde{\Pi}^{aI} n^{J} \partial_t \omega_{aIJ} &=& -2\tilde{\Pi}^{aI} n^{J} \partial_t \left ( \omega_{aIJ} - \Gamma_{aIJ} + \Gamma_{aIJ} \right ) \nonumber\\ 
			&=& -2\tilde{\Pi}^{aI} n^{J} \partial_t \left ( \omega_{aIJ} - \Gamma_{aIJ} \right )\nonumber\\
			&&  -2 \partial_a \left ( n_I \partial_t \tilde{\Pi}^{aI} \right ) \nonumber\\
		&=&	2 \tilde{\Pi}^{a I}\partial_{t}  \left [ W_{a}{}^{b}{}_{I J K} \left ( \omega_{b}{}^{J K} - \Gamma_{b}{}^{J K} \right ) \right ] \nonumber\\
		&& -2 \partial_a \left ( n_I \partial_t \tilde{\Pi}^{aI} \right ).
	\end{eqnarray}
	The reason to keep $\Gamma_{aIJ}$ with the minus sign is because $\omega_{b}{}^{JK} - \Gamma_{b}{}^{JK}$ is an
	 $SO(n-1,1)$ [or $SO(n)$] vector. The next step is to define the expression inside the brackets as the configuration variables
	 \begin{eqnarray}\label{wQ}
	 Q_{aI} := W_{a}{}^{b}{}_{I J K} \left ( \omega_{b}{}^{J K} - \Gamma_{b}{}^{J K} \right ), 
	 \end{eqnarray} 
	 and so
	 \begin{eqnarray}
	 - 2 \tilde{\Pi}^{aI} n^{J} \partial_t \omega_{aIJ} = 2 \tilde{\Pi}^{a I}\partial_{t} Q_{aI} - 2 \partial_a \left ( n_I \partial_t \tilde{\Pi}^{aI} \right ).
	  \end{eqnarray}
	The following step is to solve~\eqref{wQ} for $\omega_{aIJ}$, which gives
	\begin{eqnarray}\label{GammaQ}
	\omega_{aIJ} = \Gamma_{aIJ} + M_{a}{}^{b}{}_{IJK} Q_{b}{}^{K} +\tilde{\tilde{N}}_a{}^{bcd}{}_{IJ}\uac{\uac{u}}_{bcd},
	\end{eqnarray}
	with $M_{a}{}^{b}{}_{IJK}$ and $\tilde{\tilde{N}}_a{}^{bcd}{}_{IJ}$ given by~\eqref{equation_M} and~\eqref{equation_N}, respectively; and the variables $\uac{\uac{u}}_{abc}$ satisfy $\uac{\uac{u}}_{abc}= - \uac{\uac{u}}_{acb}$ and the traceless condition $\uac{\uac{u}}_{abc} \tilde{\tilde{h}}^{ab}=0$. The cases $n=3$ (that does not involve $\uac{\uac{u}}_{abc}$) and $n\geq 4$ must be analyzed separately as we already explained. The next step is to substitute~\eqref{GammaQ} into the action~\eqref{action_in_terms_of_Pi} and then redo the analysis performed in Sec.~\ref{sec:HA}  to eliminate the auxiliary fields $\uac{\uac{u}}_{abc}$ and thus obtain~\eqref{Thiemann_action}. This is done as follows. Substituting~\eqref{GammaQ} into~\eqref{action_in_terms_of_Pi}, we get
\begin{eqnarray}\label{action_in_terms_of_u}
	S &=&\kappa \int_{\mathbb{R} \times \Sigma}dt d^{n-1}x \left ( 2\tilde{\Pi}^{aI}\partial_{t} Q_{aI} + \omega_{tIJ} \tilde{\mathcal{G}}^{IJ} \right. \nonumber \\ 
	&& \left. - N^{a}\tilde{\mathcal{V}}_{a} - \underaccent{\tilde}{N} \tilde{\tilde{\mathcal{C}}} \right ),
	\end{eqnarray} 
	with
	\begin{subequations}
	\begin{eqnarray}
			\label{Gauss2Q} \tilde{\mathcal{G}}^{IJ} &=& 2 \tilde{\Pi}^{a [I} Q_{a}{}^{J]}, \\
			\label{vector2Q} \tilde{\mathcal{V}}_{a} &=&  2 \left( 2 \tilde{\Pi}^{b I}\partial_{[a} Q_{b] I} - Q_{a I}\partial_{b}\tilde{\Pi}^{b I} \right) \nonumber\\
			&& + \tilde{\mathcal{G}}_{I J}\left( \Gamma_a{}^{IJ} + M_{a}{}^{bIJK} Q_{bK} +\tilde{\tilde{N}}_a{}^{bcdIJ}  \uac{\uac{u}}_{bcd} \right),\notag \\
			&&\\
			\label{scalar2Q} \tilde{\tilde{\mathcal{C}}} &=& - \sigma \tilde{\Pi}^{a I} \tilde{\Pi}^{b J} R_{a b I J} + 2 \tilde{\Pi}^{a [I}\tilde{\Pi}^{|b|J]} Q_{aI} Q_{bJ}\nonumber\\
			&& + 2\sigma \Lambda h^{\frac{1}{(n-2)}} + 2 \tilde{\Pi}^{a I} n^{J} \nabla_{a}\tilde{\mathcal{G}}_{I J} \nonumber \\
			&& - \frac{(n-3)}{(n-2)} \sigma n^{I}\tilde{\mathcal{G}}^{J}{}_{K} n^{K}\tilde{\mathcal{G}}_{I J} \nonumber\\
			&& + \sigma \tilde{\tilde{h}}^{db} \tilde{\tilde{h}}^{cf} \tilde{\tilde{h}}^{e a} \uac{\uac{u}}_{abc}  \uac{\uac{u}}_{dfe}. 
	\end{eqnarray}
	\end{subequations}
	Factoring out $\tilde{\mathcal{G}}^{IJ}$ in $\tilde{\mathcal{V}}_{a}$ and $\tilde{\tilde{\mathcal{C}}}$, we obtain
	\begin{eqnarray}\label{action_with_lambda}
			S &=& \kappa \int_{\mathbb{R} \times \Sigma}dt d^{n-1}x\left ( 2\tilde{\Pi}^{a I}\partial_{t} Q_{a I} -\lambda_{IJ} \tilde{\mathcal{G}}^{I J} \right. \nonumber \\ 
			&& \left. -2N^{a}\tilde{\mathcal{D}}_{a} - \underaccent{\tilde}{N} \tilde{\tilde{\mathcal{S}}} \right ),
	\end{eqnarray}
	with
	\begin{subequations}
	\begin{eqnarray}
	\label{Gauss3Q} \tilde{\mathcal{G}}^{I J} &=& 2 \tilde{\Pi}^{a [I} Q_{a}{}^{J]}, \\
	\label{diffeoQ} \tilde{\mathcal{D}}_{a} &:=& 2\tilde{\Pi}^{bI} \partial_{[a} Q_{b] I} - Q_{aI}  \partial_{b}\tilde{\Pi}^{bI}, \\ 
	\label{scalar4Q} \tilde{\tilde{\mathcal{S}}} &:=& -\sigma \tilde{\Pi}^{a I}\tilde{\Pi}^{b J}R_{a b I J} + 2\tilde{\Pi}^{a [I}\tilde{\Pi}^{|b|J]} Q_{a I} Q_{b J} \nonumber \\
	&&  + 2\sigma h^{\frac{1}{n-2}} \Lambda + \sigma \tilde{\tilde{h}}^{db} \tilde{\tilde{h}}^{cf} \tilde{\tilde{h}}^{ea} \uac{\uac{u}}_{abc} \uac{\uac{u}}_{dfe},	
	\end{eqnarray}
	\end{subequations}
	and where we have also replaced $\omega_{tIJ}$ with $\lambda_{IJ}$ through
	\begin{eqnarray}\label{lambda}
		\omega_{tIJ} &=& -\lambda_{IJ} + N^a \left( \Gamma_{aIJ} + M_a{}^b{}_ {IJK} Q_b{}^K +  \tilde{\tilde{N}}_a{}^{bcd}{}_{IJ}\uac{\uac{u}}_{bcd} \right) \nonumber\\
		&& - 2  \tilde{\Pi}^{a}{}_{[I} n_{J]} \nabla_a \underaccent{\tilde}{N} - \sigma\frac{(n-3)}{(n-2)} \underaccent{\tilde}{N} n_{[I} \tilde{\mathcal{G}}_{J]K} n^K.
	\end{eqnarray}
The action~\eqref{action_with_lambda} depends on the phase space variables $(Q_{aI}, \tilde{\Pi}^{aI})$, the Lagrange multipliers $(\lambda_{IJ}, N^a, \underaccent{\tilde}{N})$, and the auxiliary fields $\uac{\uac{u}}_{abc}$. Now, we can get rid of the variables $\uac{\uac{u}}_{abc}$ by using their own equation of motion, which is given by
\begin{eqnarray}\label{auxu}
\underaccent{\tilde}{N} \tilde{\tilde{h}}^{d[b} \tilde{\tilde{h}}^{c]e} \tilde{\tilde{h}}^{af} \uac{\uac{u}}_{dfe} =0.
\end{eqnarray}
Given that $\underaccent{\tilde}{N} \neq 0$, its solution for $\uac{\uac{u}}_{abc}$ is
\begin{eqnarray}\label{u_sol}
\uac{\uac{u}}_{abc} =0.
\end{eqnarray}
Substituting this into the constraints of~\eqref{action_with_lambda} we get precisely the Hamiltonian formulation~\eqref{Thiemann_action}.

	\section{Time Gauge}\label{TG}
	We shall fix the boost freedom to reduce  the gauge group $SO(n-1,1)$ [or $SO(n)$] to the rotation group $SO(n-1)$. This is achieved by imposing by hand the gauge condition $\tilde{\Pi}^{a0} \approx 0$, which forms a second-class set~\cite{dirac1964lectures} with the boost constraint $\tilde{\mathcal{G}}^{0 i} \approx 0$ because
	\begin{equation}
			\label{Poisson_bracket_Gauss_vs_momentum}
			\left \lbrace \tilde{\Pi}^{a 0}(t,x), \tilde{\mathcal{G}}^{0i} (t,y) \right \rbrace = \frac{\sigma }{2\kappa} \tilde{\Pi}^{a i}\delta^{n-1}(x,y)
	\end{equation}
defines an invertible $(n-1) \times (n-1)$ matrix for nondegenerate $\tilde{\Pi}^{a i}$, something that we assume. This assumption combined with $\tilde{\Pi}^{a0} \approx 0$ in turn implies $n^{i} \approx 0$. So, making the second-class 
constraints strongly equal to zero, we get from~\eqref{Gauss2}
	\begin{eqnarray}
			\label{solution_time_gauge_q} {\cal{Q}}_{a0} &=& -n_{0}\tilde{\Pi}^{b i}\partial_{b}\underaccent{\tilde}{\Pi}_{a i}, 
	\end{eqnarray}
	where $\underaccent{\tilde}{\Pi}_{a i}$ denotes the inverse of $\tilde{\Pi}^{a i}$ [we also recall that~\eqref{solution_of_Gamma_in_terms_of_Pi} implies $\Gamma_{a0i}=0$, whereas $\Gamma_{aij}$ is a function of $\tilde{\Pi}^{a i}$ and their derivatives]. So, the action~\eqref{final_action} becomes
	\begin{eqnarray}\label{action_time_gauge}
	&& S =\kappa \int_{\mathbb{R} \times \Sigma}dt d^{n-1}x \left( 2\tilde{\Pi}^{a i}\partial_{t} {\cal{Q}}_{a i} -\Lambda_{i j}\tilde{\mathcal{G}}^{i j} \right. \nonumber \\ 
			&& \left. -2N^{a}\tilde{\mathcal{D}}_{a} -  \underaccent{\tilde}{N} \tilde{\tilde{\mathcal{H}}} \right) , 
	\end{eqnarray}
with
\begin{subequations}
	\begin{eqnarray}
	\label{Gauss_time_gauge}\tilde{\mathcal{G}}^{i j} &=& 2 \tilde{\Pi}^{a [i} {\cal{Q}}_{a}{}^{j]}, \\
	\label{diffeomorphism_time_gauge}\tilde{\mathcal{D}}_{a}&=& 2\tilde{\Pi}^{b i}\partial_{[a} {\cal{Q}}_{b] i} - {\cal{Q}}_{a i}\partial_{b}\tilde{\Pi}^{b i}, \\
	\label{scalar_time_gauge}\tilde{\tilde{\mathcal{H}}} &=& - \sigma \tilde{\Pi}^{a i}\tilde{\Pi}^{b j} R_{a b i j} +  2 \tilde{\Pi}^{a [i}\tilde{\Pi}^{|b|j]} {\cal{Q}}_{a i} {\cal{Q}}_{b j} \nonumber\\
	&& + 2 \sigma [\det({\tilde{\Pi}^{ai})}]^{\frac{2}{n-2}}  \Lambda. 
	\end{eqnarray}
	\end{subequations}
In analogy with the 4-dimensional case~\cite{ashtekar1991lectures}, this formulation could be called the $SO(n-1)$ ADM formulation of general relativity~\cite{ThieBook}.
On the other hand, if the gauge fixing is imposed directly in the action~\eqref{Thiemann_action}, we have $Q_{a0}=0$ and we get exactly the action~\eqref{action_time_gauge} with $Q_{ai}$ taking the place of ${\cal{Q}}_{ai}$. The fact that $Q_{ai}= {\cal{Q}}_{ai}$ can be easily seen from the relation~\eqref{canonical_transformation}. Therefore, in the time gauge, the same formulation~\eqref{action_time_gauge} arises from both~\eqref{final_action} and~\eqref{Thiemann_action}.

\section{Conclusions}\label{Concl}
In this paper we performed, in an $SO(n-1,1)$ [or $SO(n)$] covariant fashion, the canonical analysis of the $n$-dimensional Palatini action with or without a cosmological constant~\eqref{Palatini}. We followed an strategy akin to that used in Ref.~\cite{PhysRevDTBP}, where the introduction of second-class constraints in the canonical analysis of the Holst action was entirely avoided. To that end, we expressed the components of the connection $\omega_{aIJ}$ in terms of the variables ${\cal{Q}}_{aI}$ and $\uac{\uac{\lambda}}_{abc}$ as shown in the relation~\eqref{solution_omega}. The construction underlying these variables is laid out in Sec. \ref{sec:HA}, which entails a reduction of the presymplectic structure of the theory to a canonical symplectic structure. It turns out that the variables ${\cal{Q}}_{aI}$ play the role of the configuration variables of the resulting theory, whereas the variables $\uac{\uac{\lambda}}_{abc}$ are auxiliary fields that can be eliminated from the action by using their own dynamics. The final phase space is thus parametrized by the canonical pair $({\cal{Q}}_{aI}, \tilde{\Pi}^{aI})$, where $\tilde{\Pi}^{aI}$ is related to the spatial components of the orthonormal frame by the expression~\eqref{momentum}, subject to the Gauss, diffeomorphism, and scalar constraints  \eqref{Gauss4}--\eqref{scalar3}, which are first-class and make up the full set of constraints of the theory. Therefore, the introduction of second-class constraints and the subsequent elimination of them is completely bypassed in our approach.

In addition, we have also performed the canonical transformation \eqref{canonical_transformation}, which maps $({\cal{Q}}_{aI}, \tilde{\Pi}^{aI})$ into $(Q_{aI}, \tilde{\Pi}^{aI})$; in terms of these variables, the diffeomorphism constraint remains the same, whereas the Gauss and scalar constraints get much simpler [see the expressions~\eqref{GaussQ}-\eqref{scalarQ}]. The ensuing canonical formulation \eqref{Thiemann_action} is actually the one obtained in Refs.~\cite{Bodendorfer_2013,Thiemann_2013} for the higher-dimensional Palatini action after eliminating the second-class constraints arising in the canonical analysis carried out by the authors. This procedure is long and highly nontrivial, since the resulting second-class constraints are not independent (and thus reducible) for $n>4$. In contrast, our approach is quite straightforward and leads to the Hamiltonian action~\eqref{Thiemann_action} in no time. For the sake of completeness, we detail the case $n=3$ (where there are no variables $\uac{\uac{\lambda}}_{abc}$) in Appendix \ref{AppendixA}, and also present an alternative approach for the case $n=4$ in Appendix \ref{AppendixB}. Finally, we imposed the time gauge on both actions \eqref{final_action} and \eqref{Thiemann_action}, and obtained as a result the $SO(n-1)$ ADM formulation of general relativity embodied in the action \eqref{action_time_gauge}.

It is worth stressing the simplicity and tidiness of our approach to arrive at the Hamiltonian action~\eqref{final_action}. What is really remarkable is that such a decomposition \eqref{solution_omega} of the connection exists for general relativity in all dimensions $n\geq3$ (recall that in $n=3$ there are no variables $\uac{\uac{\lambda}}_{abc}$), something that enormously simplifies the canonical analysis of the theory, as we have shown in this paper. This decomposition is not only convenient for pure gravity, but can also be employed to build up the Hamiltonian formulation of general relativity coupled to matter fields. Perhaps the most interesting case would be the coupling of a spin 1/2 field, because given that it couples directly to the $SO(n-1,1)$ connection, then the variables $\uac{\uac{\lambda}}_{abc}$ are expected to get nontrivial contributions from this matter field. On the other hand, given that the diffeomorphism and scalar constraints can be combined into a single constraint $\tilde{\mathcal{H}}_I:= h^{-1/[2(n-2)]}\bigl(2 \tilde{\Pi}^a{}_I\tilde{\mathcal{D}}_a+\sigma n_I\tilde{\tilde{\mathcal{H}}}\bigr)$, it would be really interesting to investigate how this covariant constraint is related to the Lagrangian gauge symmetry unveiled in Ref.~\cite{Montesinos1709} for the $n$-dimensional Palatini action. We finally remark that the approach of this paper can also be used to do deal with the so-called ``space gauge'' following the same ideas of Ref.~\cite{MontRomEscCel}.

\acknowledgments

This work was partially supported by Fondo SEP-Cinvestav and by Consejo Nacional de Ciencia y Tecnolog\'{i}a (CONACyT), M\'{e}xico, Grant No.~A1-S-7701. Mariano Celada gratefully acknowledges the support of a DGAPA-UNAM postdoctoral fellowship.

\appendix
\section{\label{AppendixA}Canonical analysis for $n=3$}

To perform the canonical analysis for $3$-dimensional general relativity with a cosmological constant, we start from the definition~\eqref{q_equation}, which defines a system of $6$ linear equations for the unknowns $\omega_a{}^I{}_J$ whose solution is 
\begin{equation}
			\label{solution_omega_3_dimensions}
			\omega_{aIJ}= M_{a}{}^{b}{}_{IJK} {\cal{Q}}_{b}{}^{K},			
	\end{equation}
	with
	\begin{equation}
			\label{M_3_dimensions}
			M_{a}{}^{b}{}_{I J K}= 2\sigma \left ( \delta^{b}_{a}n_{[I}\eta_{J] K} +  \uac{\uac{h}}_{a c}\tilde{\Pi}^{c}{}_{[I}\tilde{\Pi}^{b}{}_{J]}n_{K} \right ).
	\end{equation}
Notice that there are no $\uac{\uac{\lambda}}_{abc}$ variables involved.  Substituting~\eqref{solution_omega_3_dimensions} into the action~\eqref{action_in_terms_of_Pi}, we obtain
	\begin{eqnarray}
			\label{action_3_dim_q1}
			S &=&\kappa \int_{\mathbb{R} \times \Sigma}dt d^{2}x \left ( 2\tilde{\Pi}^{a I}\partial_{t} {\cal{Q}}_{a I} + \omega_{t I J}\tilde{\mathcal{G}}^{I J} \right. \nonumber \\ 
			&& -\left. N^{a}\tilde{\mathcal{V}}_{a} - \underaccent{\tilde}{N} \tilde{\tilde{\mathcal{C}}} \right ),
	\end{eqnarray}
where
	\begin{subequations}
		\begin{eqnarray}
			\label{Gauss_3_dim_q} \tilde{\mathcal{G}}^{I J} &=& 2 \tilde{\Pi}^{a [I} {\cal{Q}}_{a}{}^{J]} + 4 \delta^{I}_{[K}\delta^{J}_{L]} \tilde{\Pi}^{a [K} n^{M]} \Gamma_{a}{}^{L}{}_{M}, \\
			\label{Vector_3_dim_q} \tilde{\mathcal{V}}_{a} &=& 2 \left( 2 \tilde{\Pi}^{b I}\partial_{[a} {\cal{Q}}_{b] I} - {\cal{Q}}_{a I}\partial_{b}\tilde{\Pi}^{b I} \right) + \tilde{\mathcal{G}}_{I J} M_{a}{}^{b I J K} {\cal{Q}}_{b K}, \notag \\
			&& \\
			\label{scalar_3_dim_q} \tilde{\tilde{\mathcal{C}}} &=& - \sigma \tilde{\Pi}^{a I} \tilde{\Pi}^{b J} R_{a b I J} + 2 \tilde{\Pi}^{a [I}\tilde{\Pi}^{|b|J]} \left ( {\cal{Q}}_{a I} {\cal{Q}}_{b J}  \right. \nonumber \\
			&& \left. + 2 {\cal{Q}}_{a I}\Gamma_{b J K} n^{K} + \Gamma_{a I L}\Gamma_{b J K}n^{K}n^{L} \right ) + 2\sigma \Lambda h \notag \\
			&&+ 2 \tilde{\Pi}^{a I} n^{J} \nabla_{a}\tilde{\mathcal{G}}_{I J}.
		\end{eqnarray}
	\end{subequations}
Factoring out $\tilde{\mathcal{G}}^{I J}$ in $\tilde{\mathcal{V}}_{a}$ and $ \tilde{\tilde{\mathcal{C}}} $, we arrive at the Hamiltonian formulation of the $3$-dimensional Palatini action with a cosmological constant
\begin{eqnarray}
			\label{action_3_dim_q2}
			 S &=& \kappa \int_{\mathbb{R} \times \Sigma}dt d^{2}x \left ( 2 \tilde{\Pi}^{a I}\partial_{t} {\cal{Q}}_{a I} -\Lambda_{I J}\tilde{\mathcal{G}}^{I J} \right. \nonumber \\ 
			&& \left. - 2 N^{a}\tilde{\mathcal{D}}_{a} -  \underaccent{\tilde}{N} \tilde{\tilde{\mathcal{H}}} \right ),
	\end{eqnarray}
where
	\begin{subequations}
	\begin{eqnarray} 
	 \tilde{\mathcal{G}}^{I J} &=& 2 \tilde{\Pi}^{a [I} {\cal{Q}}_{a}{}^{J]} + 4 \delta^{I}_{[K}\delta^{J}_{L]} \tilde{\Pi}^{a [K} n^{M]} \Gamma_{a}{}^{L}{}_{M}, \\
	\label{diffeomorphism3dim_q} \tilde{\mathcal{D}}_{a} &:=& 2\tilde{\Pi}^{b I} \partial_{[a} {\cal{Q}}_{b] I} - {\cal{Q}}_{a}{}^{I}\partial_{b}\tilde{\Pi}^{b}{}_{I}, \\ 
	\label{scalar3dim_q} \tilde{\tilde{\mathcal{H}}} &:=& - \sigma \tilde{\Pi}^{a I}\tilde{\Pi}^{b J} R_{a b I J} +  2 \tilde{\Pi}^{a [I}\tilde{\Pi}^{|b|J]} \left ( {\cal{Q}}_{a I} {\cal{Q}}_{b J} \right. \nonumber \\
	&& \left. +2 {\cal{Q}}_{a I}\Gamma_{b J K}n^{K} + \Gamma_{a I K}\Gamma_{b J L}n^{K}n^{L} \right ) + 2 \sigma h\Lambda, \notag \\
	\end{eqnarray}
	\end{subequations}
are the $SO(2,1)$ [or $SO(3)$] Gauss, diffeomorphism and scalar constraints, respectively; and where we have redefined the Lagrange multiplier $\omega_{tIJ}$ through
\begin{eqnarray}
\omega_{tIJ} &=& -\Lambda_{IJ} + N^a M_a{}^b{}_{IJK} {\cal{Q}}_b{}^K - 2  \tilde{\Pi}^{a}{}_{[I} n_{J]} \nabla_a \underaccent{\tilde}{N}. \notag \\
\end{eqnarray}
It is worth mentioning that the action~\eqref{action_3_dim_q2} is precisely the {\it same} Hamiltonian formulation~\eqref{final_action} obtained in Sec.~\ref{sec:HA} for $n>3$ (when the auxiliary fields $\uac{\uac{\lambda}}_{abc}$ are present). Therefore, the Hamiltonian formulation~\eqref{final_action} holds for $n \geq 3$. 

\subsection{Canonical transformations}
To close this appendix, we perform a canonical transformation--depending on two real parameters $\alpha$ and $\beta$--that leave the momentum $\tilde{\Pi}^{a I}$ unchanged. The transformation from $({\cal{Q}}_{aI}, \tilde{\Pi}^{a I})$ to the phase space variables $(Y_{a I}, \tilde{\Pi}^{a I})$ is such that the configuration variables $Y_{aI}$ are defined by 
	\begin{equation}
			\label{transformation_Y}
			Y_{aI} := {\cal{Q}}_{aI} - \left (\alpha W_{a}{}^{b}{}_{IJK} + \frac{\sigma\beta}{2} \delta^b_a \epsilon_{IJK} \right )  \Gamma_b{}^{JK},
	\end{equation}
where $ W_{a}{}^{b}{}_{IJK}$ has been defined in~\eqref{equation_for_W}. This transformation is indeed canonical because
\begin{eqnarray}
2\tilde{\Pi}^{aI} \partial_{t} Y_{aI} &=& 2\tilde{\Pi}^{a I} \partial_{t} {\cal{Q}}_{aI} \nonumber\\
&& + \partial_a \left [ -2 \sigma \beta \epsilon^{IJK} \tilde{\Pi}^a{}_J \tilde{\Pi}^b{}_K \partial_t \left ( \uac{\uac{h}}_{bc} \tilde{\Pi}^{c}{}_I \right) \right. \nonumber\\
&& \left. + 2 \alpha n_I \partial_t \tilde{\Pi}^{aI}  \right ].
\end{eqnarray}	
Hence, in terms of the canonical variables $(Y_{a I}, \tilde{\Pi}^{a I})$, the action~\eqref{action_3_dim_q2} becomes
\begin{eqnarray}
			\label{action_3_dim_Y}
			 S &=& \kappa \int_{\mathbb{R} \times \Sigma}dt d^{2}x \left ( 2 \tilde{\Pi}^{a I}\partial_{t} Y_{a I} - 2 \sigma \Lambda_{I}\tilde{\mathcal{G}}^{I} \right. \nonumber \\ 
			&& \left. - 2 N^{a}\tilde{\mathcal{D}}_{a} - \underaccent{\tilde}{N} \tilde{\tilde{\mathcal{H}}} \right ),
	\end{eqnarray}
with
\begin{widetext}
\begin{subequations}
	\begin{eqnarray} 
	\tilde{\mathcal{G}}^{I}&:=&- \frac{1}{2} \epsilon^{I J K}\tilde{\mathcal{G}}_{J K} =  (\beta \partial_{a}\tilde{\Pi}^{aI} + \epsilon^{I}{}_{J K} Y_{a}{}^{J} \tilde{\Pi}^{aK}) - 2 (1-\alpha)\epsilon^{I}{}_{J K}\Gamma_{a}{}^{K}{}_{L}\tilde{\Pi}^{a[J}n^{L]}, \\
	 \tilde{\mathcal{D}}_{a} &=& 2\tilde{\Pi}^{bI} \partial_{[a} Y_{b]I} - Y_{a}{}^{I}\partial_{b}\tilde{\Pi}^{b}{}_{I}, \\ 
	 \tilde{\tilde{\mathcal{H}}} &=& - \sigma \tilde{\Pi}^{a I}\tilde{\Pi}^{b J}R_{a b I J} + 2 \tilde{\Pi}^{a [I}\tilde{\Pi}^{|b|J]}\left[  Y_{a I}Y_{bJ} + (1-\alpha)\Gamma_{bJK}n^{K}\bigg{(} 2Y_{a I} + \sigma\beta\epsilon_{ILM}\Gamma_{a}{}^{LM} + (1-\alpha)\Gamma_{aIL}n^{L} \bigg{)} \right. \nonumber \\
	&&  + \sigma\beta\epsilon_{IKL}\Gamma_{a}{}^{KL}Y_{bJ} + \sigma\beta^{2}\Gamma_{a}{}^{K}{}_{J}\Gamma_{bIK} \bigg{]} + 2 \sigma h\Lambda,
	\end{eqnarray}
	\end{subequations}
\end{widetext}
 and $\Lambda_I:= - \frac{1}{2} \epsilon_{IJK}\Lambda^{JK}$.

These ugly-looking expressions acquire a more familiar form for particular choices of the parameters $\alpha$ and $\beta$:
\begin{itemize}
	\item[(i)] Case $\alpha=1=\beta$. Let us denote $A_{aI} \equiv Y_{aI} \mid_{\alpha=1, \beta=1}$. Then the action~\eqref{action_3_dim_Y} takes the form
	\begin{eqnarray}\label{ashtekar3d}
		S &=& \kappa \int_{\mathbb{R} \times \Sigma}dt d^{2}x \left ( 2 \tilde{\Pi}^{a I}\partial_{t} A_{a I} - 2 \sigma \mu_{I}\tilde{\mathcal{G}}^{I} \right. \nonumber \\ 
		&& \left. - 2 N^{a}\tilde{\mathcal{D}}_{a} - \underaccent{\tilde}{N} \tilde{\tilde{\mathcal{H}}} \right ), 
	\end{eqnarray}
	with
	\begin{subequations}
		\begin{eqnarray}
			\tilde{\mathcal{G}}^{I} &=& \partial_{a}\tilde{\Pi}^{a I} + \epsilon^{I}{}_{JK} A_{a}{}^{J} \tilde{\Pi}^{aK}, \\
			\tilde{\mathcal{D}}_{a} &=& 2\tilde{\Pi}^{bI}\partial_{[a}A_{b]I} - A_{a}{}^{I}\partial_{b}\tilde{\Pi}^{b}{}_{I}, \\
			\tilde{\tilde{\mathcal{H}}} &=& \sigma \epsilon_{I J K}\tilde{\Pi}^{aI}\tilde{\Pi}^{bJ}F_{ab}{}^{K} + 2 \sigma h\Lambda,
		\end{eqnarray} 
	\end{subequations}
	where we have used the relation between the curvature $R_{ab}{}^I{}_J$ and the curvature of the $SO(2,1)$ [or $SO(3)$] connection $A_a{}^I$, $F_{ab}{}^{I} = \partial_a A_b{}^I - \partial_b A_a{}^I + \epsilon^I{}_{JK} A_a{}^J A_b{}^K$, given by 
	\begin{eqnarray}
		-&&\sigma \tilde{\Pi}^{aI}\tilde{\Pi}^{bJ}R_{abIJ} \notag\\ =&&\sigma\epsilon_{IJK}\tilde{\Pi}^{aI}\tilde{\Pi}^{bJ}F_{ab}{}^{K} 
		+ 2\sigma\tilde{\Pi}^{aI}\nabla_{a}\tilde{\mathcal{G}}_{I} \nonumber\\
		&& - 2\tilde{\Pi}^{a[I}\tilde{\Pi}^{|b|J]} \left ( A_{aI}A_{bJ} + \sigma \Gamma_{a}{}^{K}{}_{J}\Gamma_{bIK}  \right. \nonumber\\
		&& \left. + \sigma\epsilon_{IKL}\Gamma_{a}{}^{KL}A_{bJ}  \right ),
	\end{eqnarray}
	and we have also redefined the Lagrange multiplier $\Lambda_I$ as $\mu_{I}:=\Lambda_{I} -  \tilde{\Pi}^a{}_I \nabla_{a} \uac{N}$. The action~\eqref{ashtekar3d} embodies the 3-dimensional Ashtekar formalism~\cite{Peldan9400}.
	\item[(ii)] Case $\alpha=1$ and $\beta=0$. From the transformation~\eqref{transformation_Y} it is clear that $Y_{aI} \mid_{\alpha=1, \beta=0}$ becomes the $SO(2,1)$ [or $SO(3)$] vector $Q_{aI}$ given in the relation~\eqref{canonical_transformation}, i.e., $Q_{aI} = Y_{aI} \mid_{\alpha=1, \beta=0}$ and so the action~\eqref{action_3_dim_Y} takes the form~\eqref{Thiemann_action} for $n=3$ as already explained in Sec.~\ref{Qs}. 
\end{itemize}

	The relationship between $A_{aI}$ and $Q_{aI}$ is $A_{aI}=\Gamma_{aI} + Q_{aI}$, with $\Gamma_{aI}=-(\sigma/2) \epsilon_I{}^{JK} \Gamma_{aJK}$.

\section{\label{AppendixB}Alternative canonical analysis for $n=4$}
When $n=4$ the solution \eqref{solution_omega} for $\omega_{aIJ}$ can, alternatively, be expressed as
\begin{eqnarray}\label{n4_omega_alterna}
\omega_{aIJ} = M_a{}^b{}_{IJK} {\cal{Q}}_b{}^K + \tilde{N}^b{}_{IJ} \uac{\lambda}_{ab},
\end{eqnarray}
with $M_a{}^b{}_{IJK}$ still given by~\eqref{equation_M}, whereas
\begin{eqnarray}
&&\tilde{N}^a{}_{IJ}: = \epsilon_{IJKL}\tilde{\Pi}^{aK}n^{L}.
\end{eqnarray}
There are six independent variables $\uac{\lambda}_{ab}$ in~\eqref{n4_omega_alterna} because $\uac{\lambda}_{ab}= \uac{\lambda}_{ba}$. The expression~\eqref{n4_omega_alterna} comes from substituting 
\begin{eqnarray}
\uac{\uac{\lambda}}_{abc} &=& \epsilon_{IJKL} \uac{\uac{h}}_{bd} \uac{\uac{h}}_{ec} \tilde{\Pi}^{dI} \tilde{\Pi}^{eJ} \tilde{\Pi}^{fK} n^L \uac{\lambda}_{af}  \nonumber\\
&=& -\frac{\sigma}{\sqrt{h}} \uac{\eta}_{tbcd} \tilde{\tilde{h}}^{de} \uac{\lambda}_{ae} 
\end{eqnarray}
into~\eqref{solution_omega}. Notice that this expression for $\uac{\uac{\lambda}}_{abc}$ explicitly satisfies $\uac{\uac{\lambda}}_{abc}= - \uac{\uac{\lambda}}_{acb}$ and the traceless condition $\uac{\uac{\lambda}}_{abc} \tilde{\tilde{h}}^{ab}=0$. The parametrization \eqref{n4_omega_alterna} is analogous to that used in Refs.~\cite{Montesinos1801,PhysRevDTBP}.

Note that the objects $W_a{}^b{}_{IJK}$, $M_a{}^b{}_{IJK}$, $\tilde{N}^b{}_{IJ}$, and
	\begin{equation}
			\uac{U}_{ab}{}^{cIJ}:= \frac{1}{2}\epsilon^{IJKL} \delta^{c}{}_{(a} \uac{\uac{h}}_{b)e} \tilde{\Pi}^{e}{}_{K}n_{L}, 
	\end{equation} 
	satisfy the orthogonality relations
	\begin{eqnarray}
		W_{a}{}^{cIMN}M_{c}{}^{b}{}_{MNJ} &=&\delta^{b}_{a}\delta^{I}_{J}, \\
		\uac{U}_{ab}{}^{cIJ}\tilde{N}^{d}{}_{IJ} &=& \delta_a{}^{(c} \delta_b{}^{d)}, \\
		W_{a}{}^{(b}{}_{IJK}\tilde{N}^{c) JK} &=& 0, \\
		\uac{U}_{ab}{}^{cIJ}M_{c}{}^{d}{}_{IJK} &=& 0.
	\end{eqnarray}

The transformation~\eqref{n4_omega_alterna}, $({\cal{Q}}_{aI},\uac{\lambda}_{ab})\longmapsto (\omega_{aIJ})$, is invertible, with inverse map 
$(\omega_{aIJ})\longmapsto({\cal{Q}}_{aI},\uac{\lambda}_{ab})$ given by~\eqref{q_equation} and 
\begin{eqnarray}
\uac{\lambda}_{ab} &=& \uac{U}_{ab}{}^{cIJ} \omega_{cIJ},
\end{eqnarray}
establishing that ${\cal{Q}}_{aI}$ and $\uac{\lambda}_{ab}$ are independent of each other. Therefore, we can replace the variables $\omega_{aIJ}$ with  $({\cal{Q}}_{aI},\uac{\lambda}_{ab})$ by substituting~\eqref{n4_omega_alterna} into the action~\eqref{action_in_terms_of_Pi}. By doing this, we get 
\begin{eqnarray}\label{n4_action_in_terms_of_lambda}
	S &=&\kappa \int_{\mathbb{R} \times \Sigma}dt d^3x \left ( 2 \tilde{\Pi}^{a I}\partial_{t} {\cal{Q}}_{a I} + \omega_{t I J}\tilde{\mathcal{G}}^{I J} \right. \nonumber \\ 
	&& \left. - N^{a}\tilde{\mathcal{V}}_{a} - \underaccent{\tilde}{N} \tilde{\tilde{\mathcal{C}}}  \right )
	\end{eqnarray} 
	with 
	\begin{widetext}
	\begin{subequations}
	\begin{eqnarray}
			\label{Gauss2n4} \tilde{\mathcal{G}}^{I J} &=& 2 \tilde{\Pi}^{a [I} {\cal{Q}}_{a}{}^{J]} + 4 \delta^{I}_{[K}\delta^{J}_{L]} \tilde{\Pi}^{a [K} n^{M]} \Gamma_{a}{}^{L}{}_{M}, \\
			\label{vector2n4} \tilde{\mathcal{V}}_{a} &=& 2 \left( 2 \tilde{\Pi}^{b I}\partial_{[a} {\cal{Q}}_{b] I} - {\cal{Q}}_{a I}\partial_{b}\tilde{\Pi}^{b I} \right) + \tilde{\mathcal{G}}_{I J}\left( M_{a}{}^{b I J K} {\cal{Q}}_{b K} +  \uac{\lambda}_{ab} \tilde{N}^{bIJ} \right),\\
			\label{scalar2n4}  \tilde{\tilde{\mathcal{C}}} &=& - \sigma \tilde{\Pi}^{a I} \tilde{\Pi}^{b J} R_{a b I J} + 2 \tilde{\Pi}^{a [I}\tilde{\Pi}^{|b|J]} \left[ {\cal{Q}}_{a I} {\cal{Q}}_{b J} + 2 {\cal{Q}}_{a I}\Gamma_{b J K} n^{K} + \Gamma_{a I L}\Gamma_{b J K}n^{K}n^{L} \right] + 2\sigma \Lambda \sqrt{h} + 2 \tilde{\Pi}^{a I} n^{J} \nabla_{a}\tilde{\mathcal{G}}_{I J} \nonumber \\
			&&   - \frac{\sigma}{2} n^{I}\tilde{\mathcal{G}}^{J}{}_{K} n^{K}\tilde{\mathcal{G}}_{I J} + \sigma G^{abcd}(\uac{\lambda}_{ab} - \uac{U}_{ab}{}^{eIJ}\Gamma_{eIJ})(\uac{\lambda}_{cd}-\uac{U}_{cd}{}^{fKL}\Gamma_{fKL}),
	\end{eqnarray}
	\end{subequations}
	\end{widetext}
	where $G^{abcd}:=\tilde{\tilde{h}}^{ab}\tilde{\tilde{h}}^{cd} - \tilde{\tilde{h}}^{(a|c}\tilde{\tilde{h}}^{|b)d}$ has weight +4. Now, factoring out the Gauss constraint $\tilde{\mathcal{G}}^{I J}$ in $\tilde{\mathcal{V}}_{a}$ and $ \tilde{\tilde{\mathcal{C}}}$, and redefining the Lagrange multiplier $\omega_{tIJ}$, the action becomes
	\begin{eqnarray}\label{n4_action_with_Lambda}
			S &=& \kappa \int_{\mathbb{R} \times \Sigma}dt d^3x\left ( 2\tilde{\Pi}^{a I}\partial_{t} {\cal{Q}}_{a I} -\Lambda_{IJ} \tilde{\mathcal{G}}^{I J} \right. \nonumber \\ 
			&& \left. - 2N^{a}\tilde{\mathcal{D}}_{a} - \underaccent{\tilde}{N} \tilde{\tilde{\mathcal{S}}} \right ),
	\end{eqnarray}
	where 
	\begin{subequations}
	\begin{eqnarray}
	\label{Gauss2n4re}  \tilde{\mathcal{G}}^{I J} &=& 2 \tilde{\Pi}^{a [I} {\cal{Q}}_{a}{}^{J]} + 4 \delta^{I}_{[K}\delta^{J}_{L]} \tilde{\Pi}^{a [K} n^{M]} \Gamma_{a}{}^{L}{}_{M}, \\
	\label{diffeomorphism_n4} \tilde{\mathcal{D}}_{a} &:=& 2\tilde{\Pi}^{b I} \partial_{[a} {\cal{Q}}_{b] I} - {\cal{Q}}_{a}{}^{I}\partial_{b}\tilde{\Pi}^{b}{}_{I}, \\ 
	\label{scalar4_n4} \tilde{\tilde{\mathcal{S}}} &:=& -\sigma \tilde{\Pi}^{a I}\tilde{\Pi}^{b J}R_{a b I J} + 2\tilde{\Pi}^{a [I}\tilde{\Pi}^{|b|J]} \bigg{[}  {\cal{Q}}_{a I} 
	{\cal{Q}}_{bJ} \nonumber \\
	&&  +2 {\cal{Q}}_{a I}\Gamma_{b J K}n^{K} + \Gamma_{a I K}\Gamma_{b J L}n^{K}n^{L} \bigg{]} + 2\sigma \sqrt{h} \Lambda \nonumber\\
	&&  + \sigma G^{abcd}(\uac{\lambda}_{ab} - \uac{U}_{ab}{}^{eIJ}\Gamma_{eIJ})(\uac{\lambda}_{cd}-\uac{U}_{cd}{}^{fKL}\Gamma_{fKL}), \nonumber\\	
	\end{eqnarray}
	\end{subequations}
	and 
\begin{eqnarray}\label{Lambda4}
\omega_{tIJ} &=& -\Lambda_{IJ} + N^a \left( M_a{}^b{}_ {IJK} {\cal{Q}}_b{}^K +  \uac{\lambda}_{ab} \tilde{N}^b{}_{IJ} \right) \nonumber\\
&& - 2  \tilde{\Pi}^{a}{}_{[I} n_{J]} \nabla_a \underaccent{\tilde}{N} - \frac{\sigma}{2} \underaccent{\tilde}{N} n_{[I} \tilde{\mathcal{G}}_{J]K} n^K.
\end{eqnarray}
Thus, the action~\eqref{n4_action_with_Lambda} depends on the Lagrange multipliers $\Lambda_{IJ}$, $N^a$, and $\underaccent{\tilde}{N} $ as well as on ${\cal{Q}}_{aI}$, $\tilde{\Pi}^{aI}$, and $\uac{\lambda}_{ab}$. As expected, the variables $\uac{\lambda}_{ab}$ are auxiliary fields that can be fixed by using their own equation of motion
\begin{equation}
	2\sigma\uac{N}G^{abcd}(\uac{\lambda}_{cd} - \uac{U}_{cd}{}^{eIJ}\Gamma_{eIJ})=0, 
	\end{equation}
which implies, since $\uac{N} \neq 0$ and $G^{abcd}$ is invertible~\cite{PhysRevDTBP}, that
	\begin{equation}
	\uac{\lambda}_{ab}=\uac{U}_{ab}{}^{cIJ}\Gamma_{cIJ}.
	\end{equation}
Substituting this back into the constraints of the action~\eqref{n4_action_with_Lambda}, we obtain precisely the canonical formulation~\eqref{final_action} for $n=4$.

\subsection{Canonical transformations}

Now, we consider a canonical transformation--depending on some parameters $\alpha$, $\beta$, and $\gamma$ (the latter corresponds to the Immirzi parameter)--that leaves the momentum variables unchanged, whereas the configuration variables are promoted to
	\begin{equation}
	X_{aI} = {\cal{Q}}_{aI} - W_{a}{}^{b}{}_{IJK}\left( \alpha\Gamma_{b}{}^{JK} + \frac{(\beta-1)}{\gamma} \ast \Gamma_{b}{}^{J K} \right), 
	\end{equation}
	where $\ast {V}_{IJ}:=(1/2)\epsilon_{IJKL}V^{KL}$. We recall that the variables $X_{aI}$ were introduced in Ref.~\cite{PhysRevDTBP}. This transformation is canonical because the symplectic term in the action~\eqref{final_action} changes by a total derivative:
	\begin{eqnarray}
	&& 2\tilde{\Pi}^{aI}\partial_{t} {\cal{Q}}_{aI}  = 2\tilde{\Pi}^{aI}\partial_{t} X_{aI}\nonumber\\
	&& +\partial_{a}\left[- 2\alpha n_I \partial_{t}\tilde{\Pi}^{aI} + \frac{\sigma(\beta-1)}{\gamma}\sqrt{h}\tilde{\eta}^{tabc}\uac{\uac{h}}_{bd}\uac{\uac{h}}_{cf}\tilde{\Pi}^{f}{}_{I}\partial_{t}\tilde{\Pi}^{dI} \right]. \nonumber\\
	\end{eqnarray}
	In terms of the new phase-space variables $(X_{aI},\tilde{\Pi}^{aI})$, the action~\eqref{final_action} for $n=4$ acquires the form
	\begin{eqnarray}
	S &=& \kappa \int_{\mathbb{R} \times \Sigma}dt d^3x \left ( 2 \tilde{\Pi}^{a I}\partial_{t} X_{aI} -\Lambda_{I J}\tilde{\mathcal{G}}^{I J} \right. \nonumber \\ 
			&& \left. - 2 N^{a}\tilde{\mathcal{D}}_{a} -  \underaccent{\tilde}{N} \tilde{\tilde{\mathcal{H}}} \right ), 
	\end{eqnarray}
	with
	\begin{widetext}
	\begin{subequations} 
	\begin{eqnarray} 
	\tilde{\mathcal{G}}^{I J} &=& 2 \tilde{\Pi}^{a[I} X_{a}{}^{J]} + 4\left[  (1-\alpha)\delta^{I}_{[K}\delta^{J}_{L]} + \frac{(1-\beta)}{2\gamma}\epsilon^{IJ}{}_{KL} \right] \tilde{\Pi}^{a[K}n^{M]}\Gamma_{a}{}^{L}{}_{M}, \\
	\tilde{\mathcal{D}}_{a} &=& 2\tilde{\Pi}^{bI}\partial_{[a} X_{b]I} - X_{aI}\partial_{b}\tilde{\Pi}^{bI}, \\
	\tilde{\tilde{\mathcal{H}}} &=& - \sigma \tilde{\Pi}^{aI}\tilde{\Pi}^{bJ}R_{abIJ} + 2 \tilde{\Pi}^{a[I}\tilde{\Pi}^{|b|J]}\bigg{\lbrace} X_{aI} X_{bJ} + \left( \frac{1-\beta}{\gamma} \right)^{2}q^{KL}\Gamma_{aIK}\Gamma_{bJL} + 2 X_{aI}\left[ (1-\alpha)\Gamma_{bJK} + \frac{(1-\beta)}{\gamma} \ast \Gamma_{bJK}\right]n^{K} \nonumber\\
	&& + (1-\alpha)\left[ (1-\alpha)\Gamma_{aIK} + \frac{2}{\gamma}(1-\beta) \ast \Gamma_{aIK} \right] \Gamma_{bJL}n^{K}n^{L} \bigg{\rbrace} + 2 \sigma\Lambda\sqrt{h}.
	\end{eqnarray}
	\end{subequations}
	
	This Hamiltonian formulation becomes more familiar for particular values of the parameters:
	\begin{itemize}
		\item[(i)] For $\alpha=1=\beta$, the configuration variable is $X_{aI}\mid_{\alpha=1,\beta=1} =Q_{aI}$, for which we recover the formulation~\eqref{Thiemann_action} for $n=4$.\\
		\item[(ii)] For $\alpha=1$ and $\beta=0$, the configuration variable is $X_{aI}\mid_{\alpha=1,\beta=0} =K_{aI}$. The action becomes
		\begin{eqnarray}
		S &=& \kappa \int_{\mathbb{R} \times \Sigma}dt d^3x \left ( 2 \tilde{\Pi}^{a I}\partial_{t} K_{a I} -\Lambda_{I J}\tilde{\mathcal{G}}^{I J}  - 2 N^{a}\tilde{\mathcal{D}}_{a} - \underaccent{\tilde}{N} \tilde{\tilde{\mathcal{H}}} \right ), 
		\end{eqnarray}
		with
		
		\begin{subequations} 
		\begin{eqnarray} 
		&& \tilde{\mathcal{G}}^{I J}=2 \tilde{\Pi}^{a[I} K_{a}{}^{J]} + \frac{2}{\gamma}\epsilon^{IJ}{}_{KL}  \tilde{\Pi}^{a[K}n^{M]}\Gamma_{a}{}^{L}{}_{M}, \\
		&&\tilde{\mathcal{D}}_{a} = 2\tilde{\Pi}^{bI}\partial_{[a} K_{b]I} - K_{aI}\partial_{b}\tilde{\Pi}^{bI}, \\
		&& \tilde{\tilde{\mathcal{H}}}= -\sigma  \tilde{\Pi}^{aI}\tilde{\Pi}^{bJ}R_{abIJ} + 2 \tilde{\Pi}^{a[I}\tilde{\Pi}^{|b|J]} \left ( K_{aI} K_{bJ} + \frac{1}{\gamma^2} q^{KL}\Gamma_{aIK}\Gamma_{bJL} + \frac{2}{\gamma} K_{aI}  \ast \Gamma_{bJK} n^{K} \right ) + 2 \sigma\Lambda\sqrt{h}.
		\end{eqnarray}
		\end{subequations}
		
		This formulation was also obtained after applying a canonical transformation on the Hamiltonian theory resulting from the Holst action~\cite{Montesinos1801}.

	\item[(iii)] For $\alpha=0=\beta$, the configuration variable is $X_{aI}\mid_{\alpha=0,\beta=0} =C_{aI}$. The action acquires the form
	\begin{eqnarray}
	S &=& \kappa \int_{\mathbb{R} \times \Sigma}dt d^3x \left ( 2 \tilde{\Pi}^{a I}\partial_{t} C_{a I} -\Lambda_{I J}\tilde{\mathcal{G}}^{I J}  - 2 N^{a}\tilde{\mathcal{D}}_{a} - \underaccent{\tilde}{N} \tilde{\tilde{\mathcal{H}}} \right ), 
	\end{eqnarray}
	with
	
	\begin{subequations} 
	\begin{eqnarray} 
	\tilde{\mathcal{G}}^{I J} &=&2 \tilde{\Pi}^{a[I} C_{a}{}^{J]} + 4\left[ \delta^{I}_{[K}\delta^{J}_{L]} + \frac{1}{2\gamma}\epsilon^{IJ}{}_{KL} \right] \tilde{\Pi}^{a[K}n^{M]}\Gamma_{a}{}^{L}{}_{M}, \\
	\tilde{\mathcal{D}}_{a} &=& 2\tilde{\Pi}^{bI}\partial_{[a} C_{b]I} - C_{aI}\partial_{b}\tilde{\Pi}^{bI}, \\
	\tilde{\tilde{\mathcal{H}}} &=& - \sigma \tilde{\Pi}^{aI}\tilde{\Pi}^{bJ}R_{abIJ} + 2 \tilde{\Pi}^{a[I}\tilde{\Pi}^{|b|J]}\bigg{\lbrace} C_{aI} C_{bJ} + \frac{1}{\gamma^2} q^{KL}\Gamma_{aIK}\Gamma_{bJL} + 2 C_{aI}\left[ \Gamma_{bJK} + \frac{1}{\gamma} \ast \Gamma_{bJK}\right]n^{K} \nonumber\\
	&& + \left[ \Gamma_{aIK} + \frac{2}{\gamma} \ast \Gamma_{aIK} \right] \Gamma_{bJL}n^{K}n^{L} \bigg{\rbrace} + 2 \sigma\Lambda\sqrt{h}.
	\end{eqnarray}
	\end{subequations}

	This Hamiltonian formulation was originally obtained in Ref.~\cite{Montesinos1801} by performing the canonical analysis of the Holst action.
	\end{itemize}
	\end{widetext}

 	\bibliographystyle{apsrev4-1}
	\bibliography{references}

\end{document}